\documentclass[%
 reprint,
 amsmath,amssymb,
 aps,
]{revtex4-1}

\usepackage{graphicx}
\usepackage{dcolumn}
\usepackage{bm}

\begin{document}


\title{Secondary electron emission and yield spectra of metals from Monte Carlo simulations and experiments}

\author{Martina Azzolini}\email{mazzolini@ectstar.eu}
\affiliation{European Centre for Theoretical Studies in Nuclear Physics and Related Areas (ECT*-FBK) and Trento Institute for Fundamental Physics and Applications (TIFPA-INFN), Trento, Italy \\
Laboratory of Bio-Inspired and Graphene Nanomechanics, Department of Civil, Environmental and Mechanical Engineering, University of Trento,Trento, Italy\\}

\author{Marco Angelucci}
\affiliation{LNF-INFN, P.O. box 13, 00044 Frascati, Roma, Italy}
\author{Rosanna Larciprete}
\affiliation{LNF-INFN, P.O. box 13, 00044 Frascati, Roma, Italy \\ CNR-ISC Institute for Complex System, via dei Taurini 19, 00185 Roma, Italy}
\author{Roberto Cimino}
\affiliation{LNF-INFN, P.O. box 13, 00044 Frascati, Roma, Italy}
\author{Nicola M. Pugno}
\affiliation{Laboratory of Bio-Inspired and Graphene Nanomechanics, Department of Civil, Environmental and Mechanical Engineering, University of Trento,Trento, Italy \\Ket-Lab, Edoardo Amaldi Foundation, Italian Space Agency, Rome, Italy \\School of Engineering and Materials Science, Queen Mary University of London, United Kingdom
}
\author{Simone Taioli}
\affiliation{European Centre for Theoretical Studies in Nuclear Physics and Related Areas (ECT*-FBK) and Trento Institute for Fundamental Physics and Applications (TIFPA-INFN), Trento, Italy \\
Faculty of Mathematics and Physics, Charles University, Prague, Czech Republic}
\author{Maurizio Dapor} \email{dapor@ectstar.eu}
\affiliation{European Centre for Theoretical Studies in Nuclear Physics and Related Areas (ECT*-FBK) and Trento Institute for Fundamental Physics and Applications (TIFPA-INFN), Trento, Italy\\}

\begin{abstract}
In this work, we present a computational method, based on the Monte Carlo statistical approach, for calculating electron energy emission and yield spectra of metals, such as copper, silver and gold. The calculation of these observables proceeds via the Mott theory to deal with the elastic scattering processes, and by using the Ritchie dielectric approach to model the electron inelastic scattering events. In the latter case, the dielectric function, which represents the starting point for the evaluation of the energy loss, is obtained from experimental reflection electron energy loss spectra. The generation of secondary electrons upon ionization of the samples is also implemented in the calculation. A remarkable agreement is obtained between both theoretical and experimental electron emission spectra and yield curves. 

\end{abstract}

\pacs{Valid PACS appear here}
\maketitle


\maketitle
\section{Introduction}
\label{introduction}

The emission of secondary electrons plays a fundamental role in materials characterization techniques, such as scanning electron microscopy \citep{seiler1983secondary, goldstein2017scanning}, and in affecting the performance of a variety of electron devices, such as the detectors based on electron multipliers \citep{sauli1997gem, benlloch1998further}. These techniques in particular seek a high value of the electron yield to reach a low noise-to-signal ratio for enhancing the image quality. High electron yield can be achieved by coating the photomultiplier with low work function photocathode or by increasing the local curvature of the surface by steep edges.\\ 
\indent On the other hand, in other applications the emission of secondary electrons must be suppressed, e.g. in particle accelerators. Indeed, detrimental effects on the machine stability, which might result in beam loss \citep{lyneis1977elimination, somersalo1998computational, cimino2004can, cimino2012nature, larciprete2013secondary, cimino2014electron}, can be caused by the so-called multipactor effect. This phenomenon appears when the current of re-emitted electrons grows uncontrollably due to presence of electronic charges in proximity of the vacuum tube walls. The latter are accelerated by the primary beam, producing an avalanche of secondary electrons. 
In this regard, chemical treatment, coating \citep{baglin2000secondary, vallgren2011amorphous} and patterning of the target surface \citep{montero2014secondary, valizadeh2014low, valizadeh2017reduction, calatroni2017first} were used to overcome the harmful effects brought about by this phenomenon. \\
\indent Despite these remarkable attempts, several issues could be bypassed by developing an efficient and accurate method to calculate the electron yield of the investigated material, in order to predict the secondary emission and thus to tailor the solution according to the application sought for. 
In this regard, analytic descriptions of the secondary electron energy yield were presented in the literature \citep{vaughan1989new, dionne1975origin, shih1997secondary, lin2005new, xie2009formula}. \\ 
\indent In this work, we present an accurate computational approach, based on the Monte Carlo method \citep{dapor2011secondary}, to simulate electron trajectories leading to secondary electron emission and we compare theoretical lineshapes with our recorded experimental yield spectra. Within this approach, elastic collision and inelastic scattering processes are carefully evaluated. In the former case this means to assess the angular deviation along the path of the electrons in their way out of the solid, while in the latter the electron energy loss. In particular, the elastic scattering is treated within the Mott theory \citep{mott1929scattering}, while the inelastic scattering events are dealt with the Ritchie dielectric theory \citep{Ritchie_PhysRev_1957}. The Monte Carlo method is used to simulate the secondary spectra of three metallic targets, notably copper, silver and gold. 
The dielectric functions used in the assessment of the energy loss of these materials are obtained from reflection electron energy loss (REEL) experiments \citep{werner2009optical}. This treatment thus takes into account the contribution to the secondary emission spectra of both bulk and surface plasmon excitations, increasing the accuracy of the computed data.  

This paper is structured as follows: in the following section II the experimental procedures used for the experimental measurements of the secondary emission and yield spectra of all metals are described. Afterwards, a detailed discussion of the computational Monte Carlo approach is presented in section III. 
Finally, in section IV we present a thorough comparison between simulated and experimental data of secondary emission and yield spectra.   

\section{Experimental details}

The experimental apparatus used to study the secondary electron yield (SEY) and angle integrated energy distribution curves (EDC) is hosted in the “Material Science” laboratory of LNF-INFN, Frascati (Rome).  For our experiments we used a specially built UHV $\mu$-metal chamber with less than 5 mGauss residual magnetic field at the sample position, pumped by a CTI8 cryo-pump to ensure a vacuum better than 10$^{-10}$  mbar. The set-up has been designed to limit the residual magnetic field near the sample, which can deviate low-energy electrons. Ion pumps are not used due to their detrimental stray magnetic field. This set up is routinely measuring SEY curves from very low primary energies to about 1000 eV \cite{cimino2004can, cimino2012nature, larciprete2013secondary, cimino2014electron, cimino2015detailed, gonzalez2017secondary}. The SEY, i. e. the ratio of the number of electrons leaving the sample surface ($I_\mathrm{s}$) to the number of incident electrons ($I_\mathrm{p}$) per unit area, was determined experimentally by measuring $I_\mathrm{p}$ and the total sample current $I_\mathrm{t}=I_\mathrm{p}-I_\mathrm{s}$, so that $\delta = 1 - I_\mathrm{t}/I_\mathrm{p}$. For the SEY measurements, the electron beam was set to be smaller than 1 mm$^2$ in transverse cross-sectional area at the sample surface. To measure the current of the impinging primary electrons, a negative bias voltage (-75 V) was applied to the sample. The SEY measurements were performed at normal incidence, by using electron beam currents of a few nA.
A SpectraLEED Omicron LEED/Auger retarding field  (RF) analyser system was specially modified to be able to collect angle integrated EDC with RF filtering and computer control while using the gun in LEED mode, i.e. with a low-energy focused beam. The e$^-$ gun provided a small and stable (both in current and position) beam spot on the sample, in the energy range from 30 to 1000 eV. This set-up, can measure angle integrated EDC with the limitation typical of any RF analyzers.  While the secondaries are consistently measured at all primary energies, elastic peaks broaden due to increasingly poor resolution at higher primary energies, showing an additional strong asymmetry on the low energy side due to the integration of the background \cite{gergely2002elastic, sulyok1992spectrometer}.  Despite those known limitations, the data can be used in this work to extract the relevant information needed.
The sample can be transferred from air into UHV conditions and can be cooled down to 10 K, exposed to various types of gasses and cleaned by subsequent cycles of Ar sputtering. For the polycrystalline Cu sample here studied, we cleaned it by repeated Ar$^+$ sputtering cycles at 1.5 KeV in Ar pressure of $5 \times 10^{-6}$ mbar until no signal of C and O was observed in the XPS spectrum.

\section{Computational details}

\subsection{Monte Carlo}

\begin{figure}[h!]
    \centering
    \includegraphics[width = 0.40\textwidth]{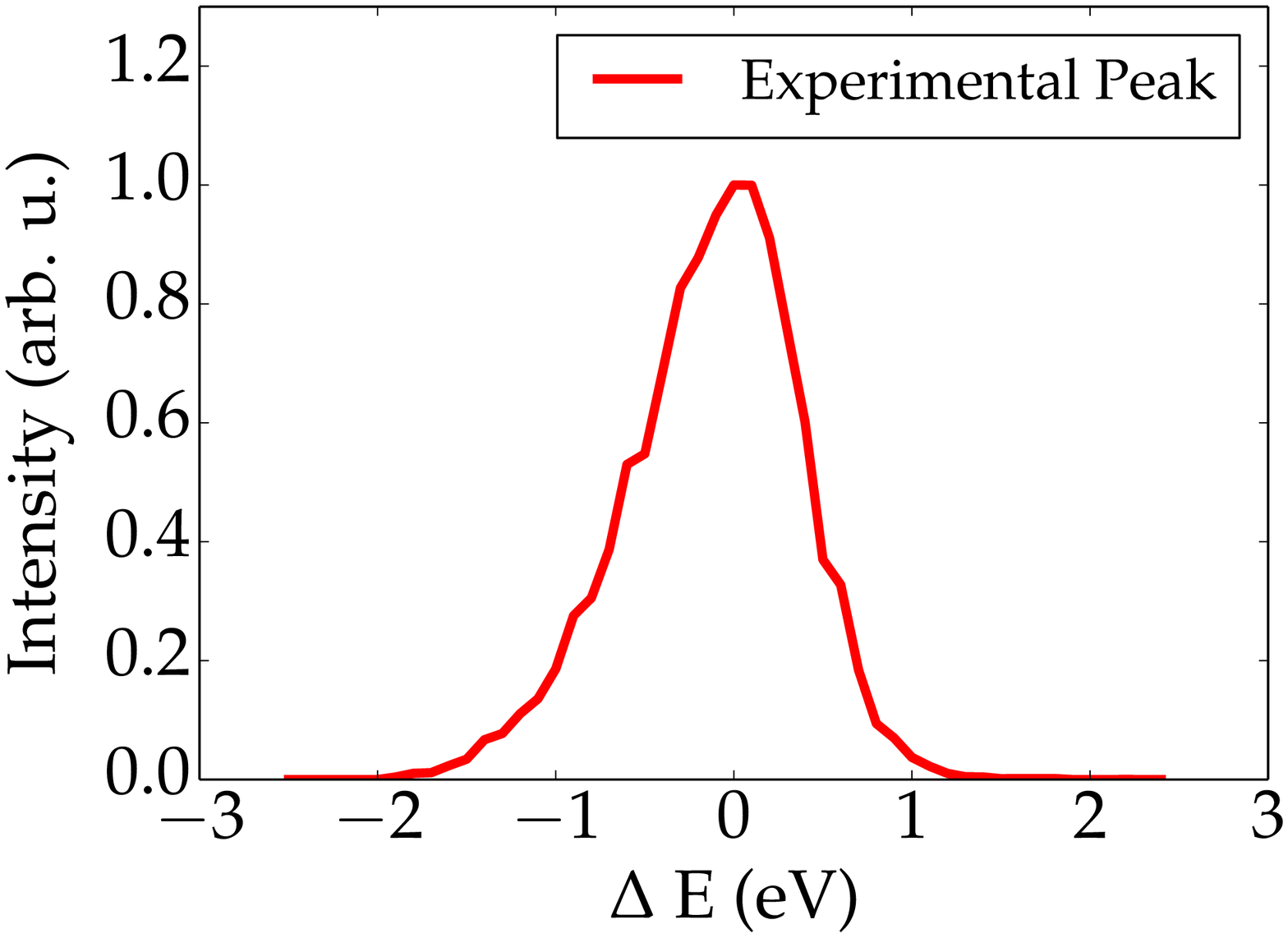}
    \includegraphics[width = 0.40\textwidth]{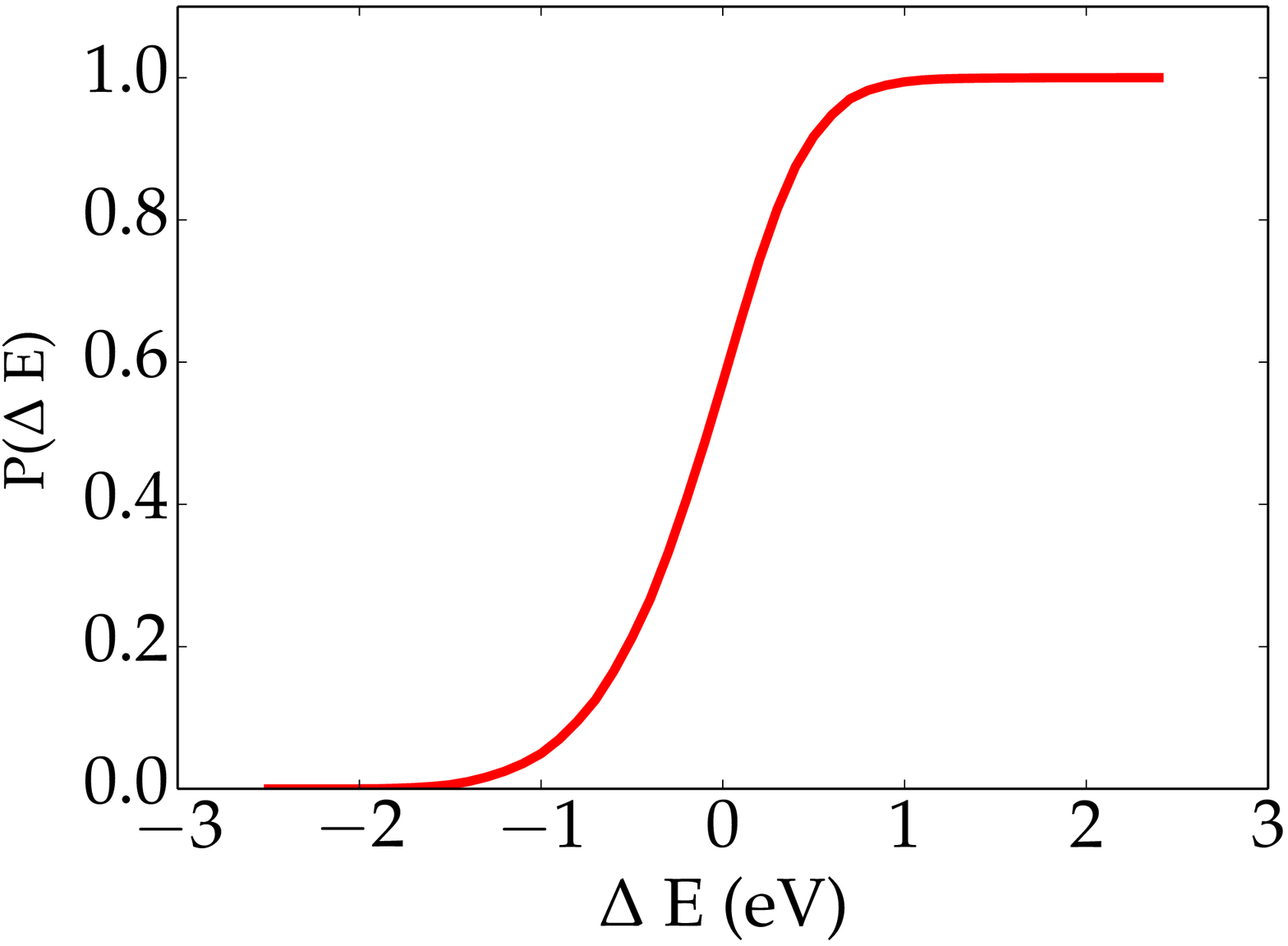}
    \caption{a) Experimental elastic peak of copper; b) Cumulative probability distribution of the experimental elastic peak.}
    \label{fig:EXP_peak}
\end{figure}

The Monte Carlo approach models the spectral distribution of energy transferred to a specimen bombarded with a perpendicularly incident electron beam by following the electron trajectories within the target. The electron trajectories result from the elastic and inelastic interactions undergone by the electrons scattered by the nuclei and the electron clouds of target atoms, respectively.\\ 
\indent Secondary electrons are generated via inelastic interactions through the ionization of atomic centers of the target. In the latter process, by measuring the kinetic energies of the escaping charges, the electron emission energy spectra can be recorded. In the Monte Carlo simulation of this emission mechanism the trajectories of secondary electrons, similarly to those generated by elastic and non-ionizing inelastic scattering, are followed by using a statistical algorithm. At variance, the path of electrons with kinetic energies below the value of the work function and of electrons emitted by the specimen are considered terminated.\\ 
\indent In Monte Carlo simulations, the probability of elastic and inelastic scattering events is assessed by comparing random numbers with the correspondent probability distributions.
These probability distributions are calculated by using the elastic and inelastic cross sections, computed by applying the Mott \citep{mott1929scattering} and the Ritchie dielectric theories \citep{Ritchie_PhysRev_1957} respectively. These models and the relevant distribution probabilities will be presented in detail in sections IIIB and IIIC. \\
\indent The starting point of charge transport Monte Carlo calculations is to determine the initial kinetic energy and the cumulative distribution probability of the primary beam impinging on the surface target. The kinetic energy can be obtained from our experiments by considering the energy distribution of the elastic peak, generated by the electrons which have been elastically reflected by the target surface. These electrons are recorded around a sharp peak at energies $\overline{E}\pm \Delta E$, where $\overline{E}$ is the energy of the peak maximum and ${\Delta E}$ takes into account the width of the kinetic energy distribution. In particular, in this work we used the experimental elastic peak distribution of copper $f(\Delta E)$, which is reported in Fig. \ref{fig:EXP_peak}a. This elastic distribution is conventionally centered in zero. Then, the cumulative distribution probability (see Fig. \ref{fig:EXP_peak}b) is computed as a function of the energy correction by using the following expression:
\begin{equation}\label{PE}
  P(\Delta E)= \frac{1}{Area}\int\limits_{E_{-}}^{\Delta E} f(\Delta E') d(\Delta E').
\end{equation}
where the total area of the peak ($Area$) was calculated by integrating the experimental $f(\Delta E)$ curve in the symmetric energy interval $[E_-; E_+]$, with $E_{\pm} = \pm 2.5$ eV as follows: 
\begin{equation}
   Area = \int\limits_{E_{-}}^{E_{+}} f(\Delta E') d(\Delta E').
\end{equation}
The value of the correction $\Delta E$ to the maximum $\overline{E}$ of the elastic peak has to be determined to set the initial kinetic energy for each electron in the primary beam. 
This is accomplished by generating a random number $\mu_1$, uniformly distributed in the interval [0,1], and by finding the value of ${\Delta E}$ for which $P(\Delta E)$ in Eq. \ref{PE} is  equal to $\mu_1$. Finally, the initial kinetic energy of each electron in the primary beam is set to $\overline{E} + {\Delta E}$ and the Monte Carlo evaluation of its trajectory within the solid resulting from elastic, inelastic and ionizing events can be pursued.\\
\indent The incident electron beam direction is set perpendicular to the specimen surface and the kinetic energy is increased by the specific target work function $\chi$. 
Upon elastic and inelastic scattering events, the electron trajectory can reach the target surface. Finally, electrons can be emitted from the sample provided that the following condition is satisfied:
\begin{equation}
E \cos^2 \theta \ge \chi,
\label{eq:emissione}
\end{equation}
where $\theta$ is the angle between the scattering direction inside the material and the normal to the target surface, and $E$ is the electron kinetic energy. This condition stems from the fact that the target-vacuum interface represents an energy barrier to be overcome by the escaping electrons at the interface. Finally, the electron emission is determined by assessing the transmission coefficient $t$, which can be obtained by assuming that the electrons feel a model step potential at the surface \citep{Dapor_book_blu}: 
\begin{equation}
    t = \frac{4\sqrt{1 - \chi/(E\cos^2 \theta)}}{\left[1 + \sqrt{1 - \chi/(E\cos^2 \theta)} \right]^2}.
\end{equation}
In our Monte Carlo approach the transmission coefficient is compared with a random number $\mu_2$, uniformly sampled in the interval [0,1] so that 
for $\mu_2 < t$ the electrons is emitted from the surface with a kinetic energy decreased by the work function $\chi$, otherwise electrons are elastically reflected back and continue their path within the solid target. 
Here we notice that $\chi$ plays a key role in the electron emission process, and we do expect a dramatic dependence of the electron yield from this observable. This effect will be discussed thoroughly in section IVA.

\subsection{Elastic scattering}

\begin{figure}[h!]
    \centering
    \includegraphics[width = 0.32\textwidth]{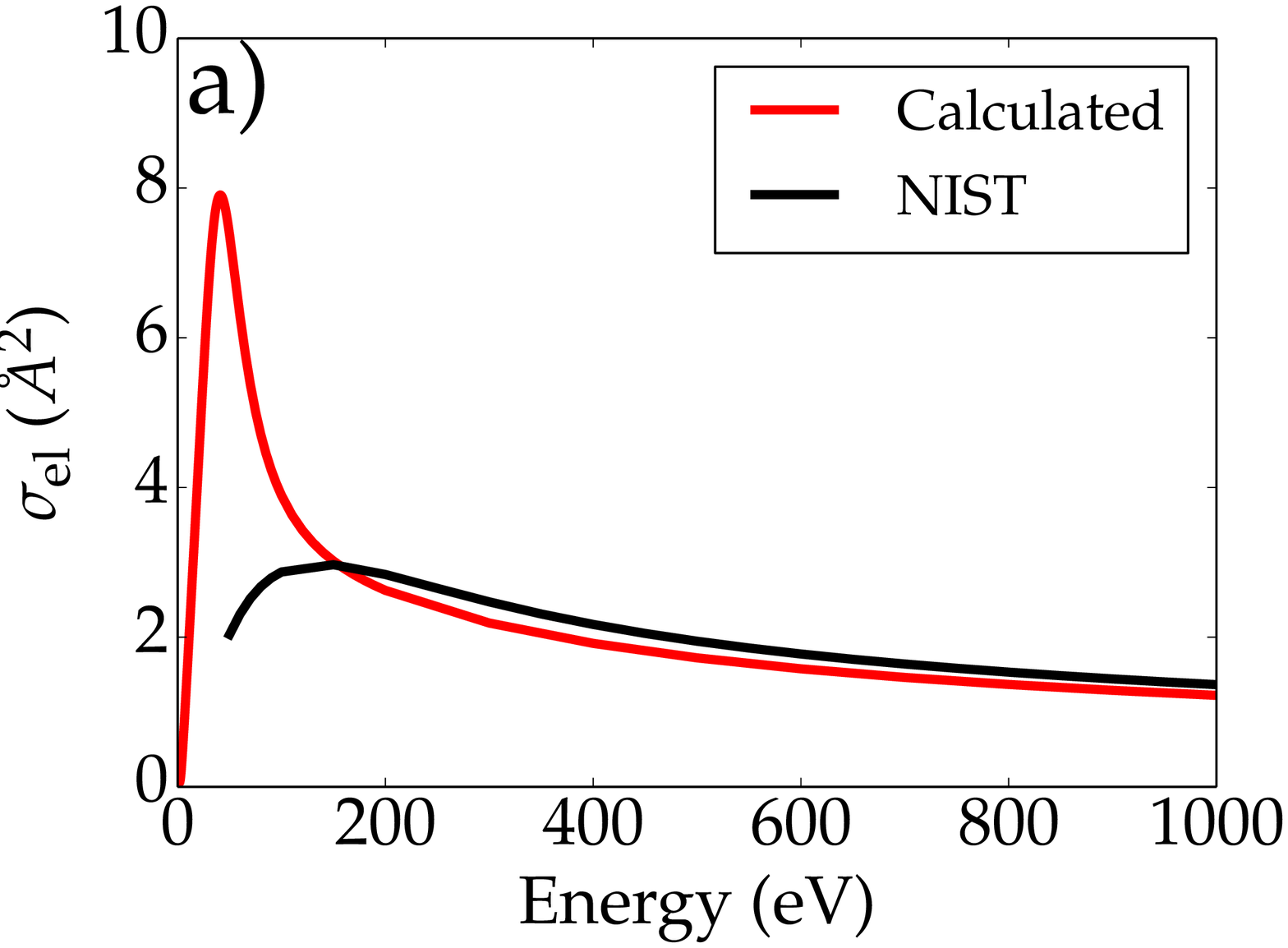}
    \includegraphics[width = 0.32\textwidth]{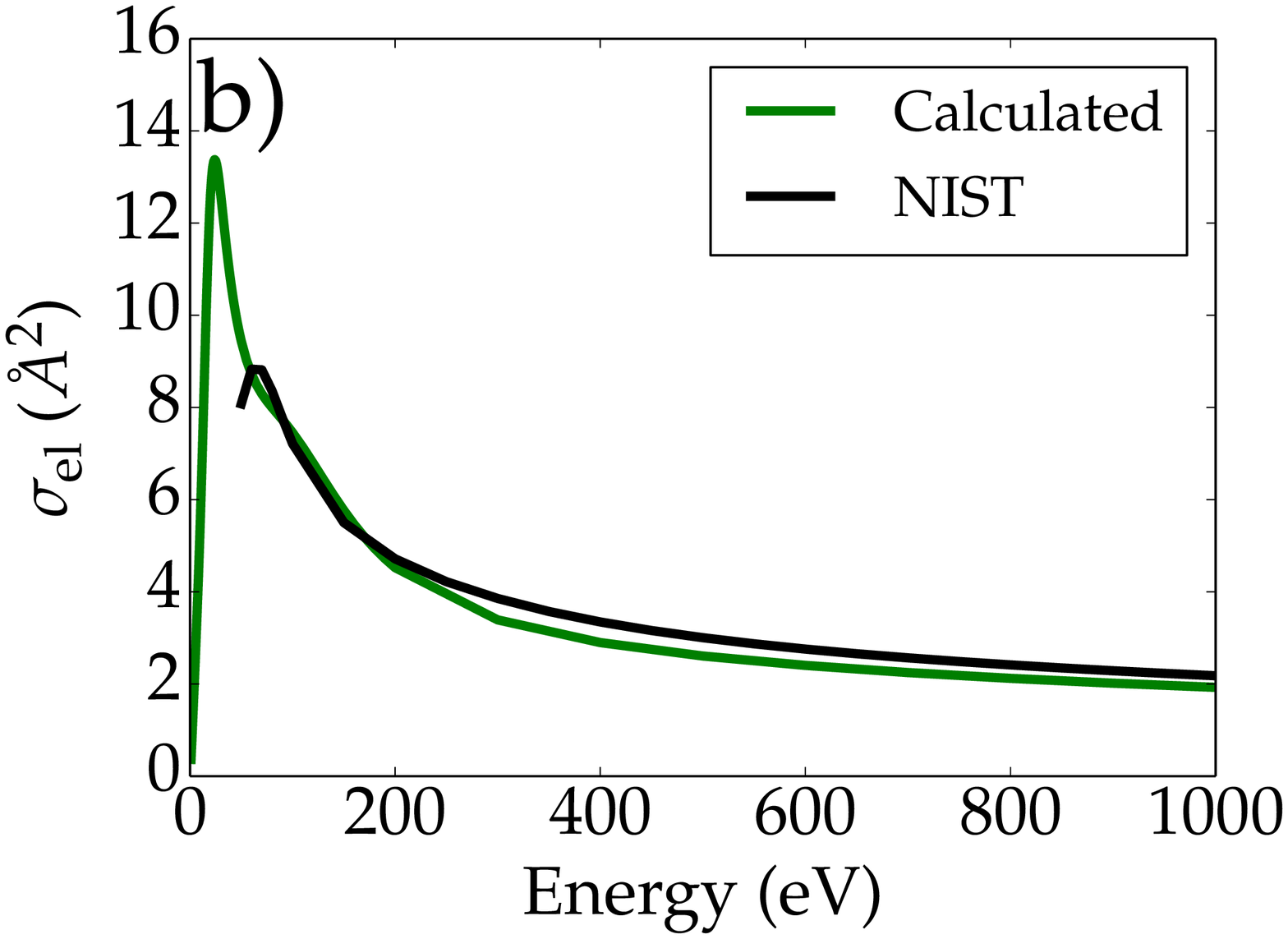}
    \includegraphics[width = 0.32\textwidth]{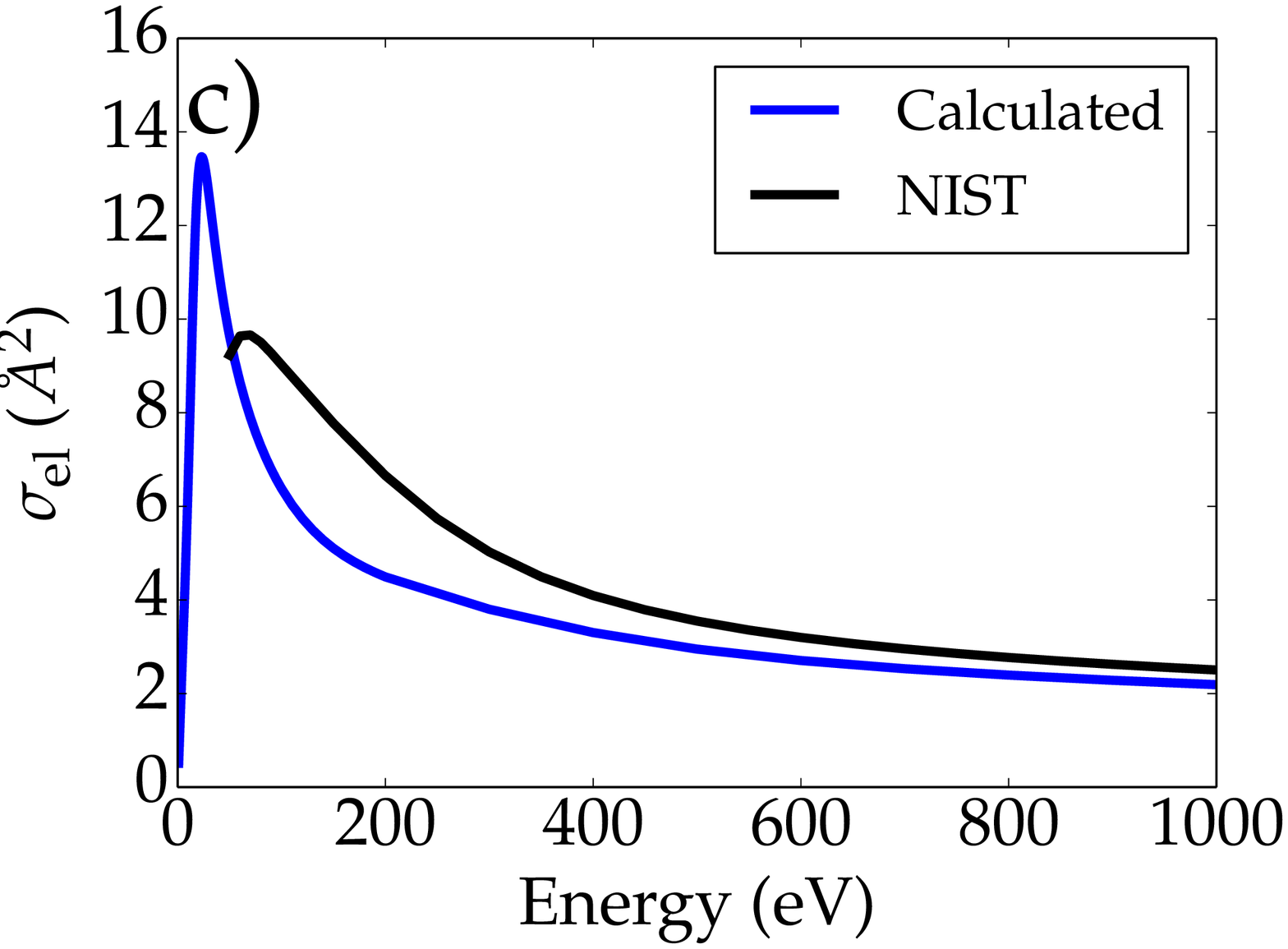}
    \caption{Total elastic scattering cross sections of a) Cu, b) Ag, and c) Au as a function of the electron kinetic energy. Black lines report tabulated values \citep{jablonski2010nist}}
    \label{fig:escs}
\end{figure}

The elastic scattering between electrons and target nuclei is described by the Mott theory \citep{mott1929scattering}. The differential elastic scattering cross section (${d\sigma_\mathrm{el}}/{d\Omega}$) can be written as:
\begin{equation}
    \frac{d\sigma_\mathrm{el}}{d\Omega} = |f|^2 + |g|^2
    \label{eq:descs}
\end{equation}
where $f$ and $g$ are the scattering amplitudes \citep{Dapor_book_blu, jablonski2004comparison}, which can be obtained by solving the Dirac equation in a central field. For the materials under investigation in this work, that is Cu, Ag and Au, the calculation of the elastic scattering cross section was performed using the analytic formulation of the atomic potential proposed by Salvat \citep{PhysRevA.36.467}. 
The total elastic scattering cross section $\sigma_{el}(E)$ is obtained by integrating Eq. (\ref{eq:descs}) in the solid angle. The results are shown in Fig. \ref{fig:escs} and compared with the tabulated values of Ref. \citep{jablonski2010nist}. 

\begin{figure}[h!]
    \centering
    \includegraphics[width = 0.32\textwidth]{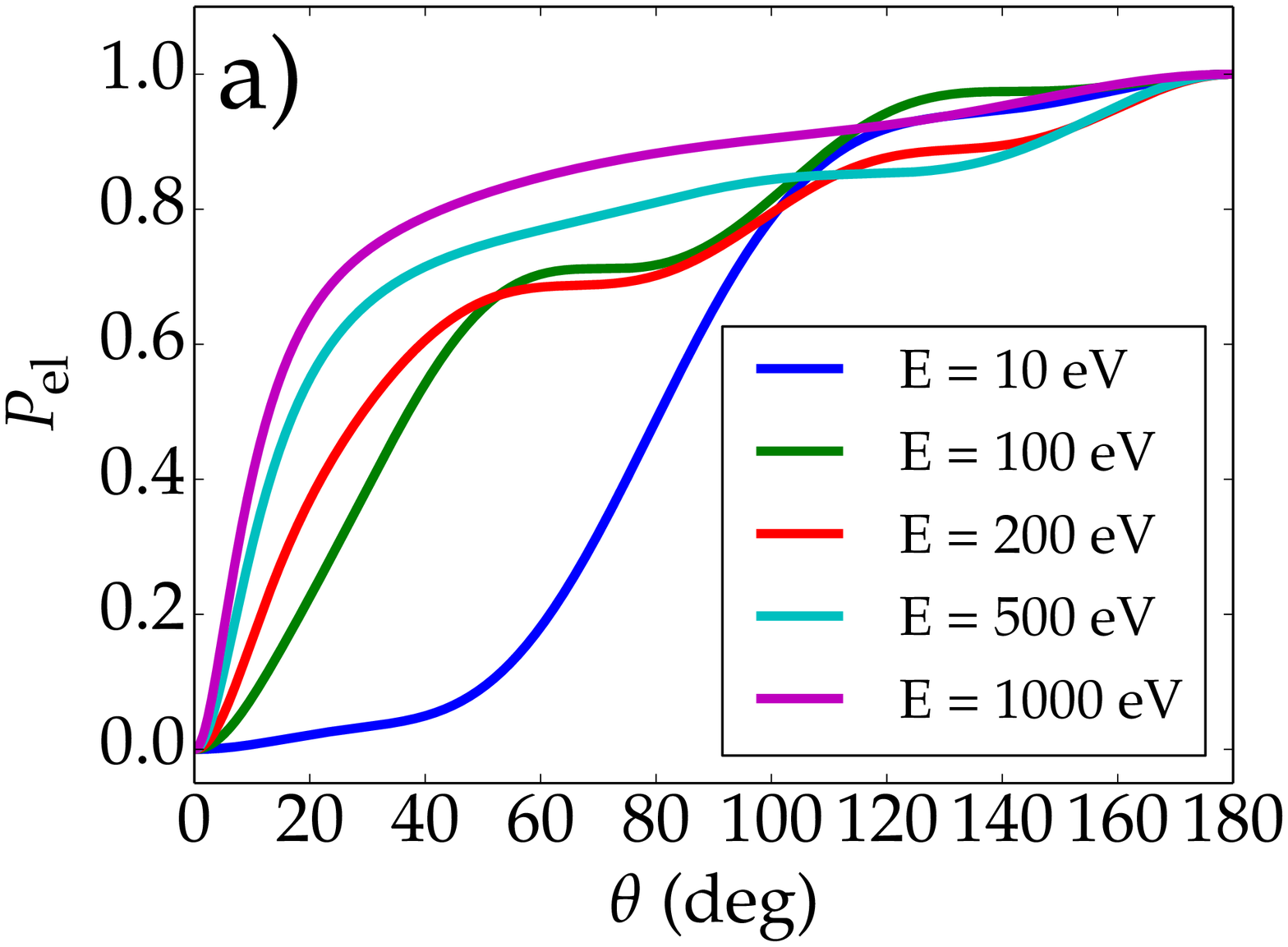}
    \includegraphics[width = 0.32\textwidth]{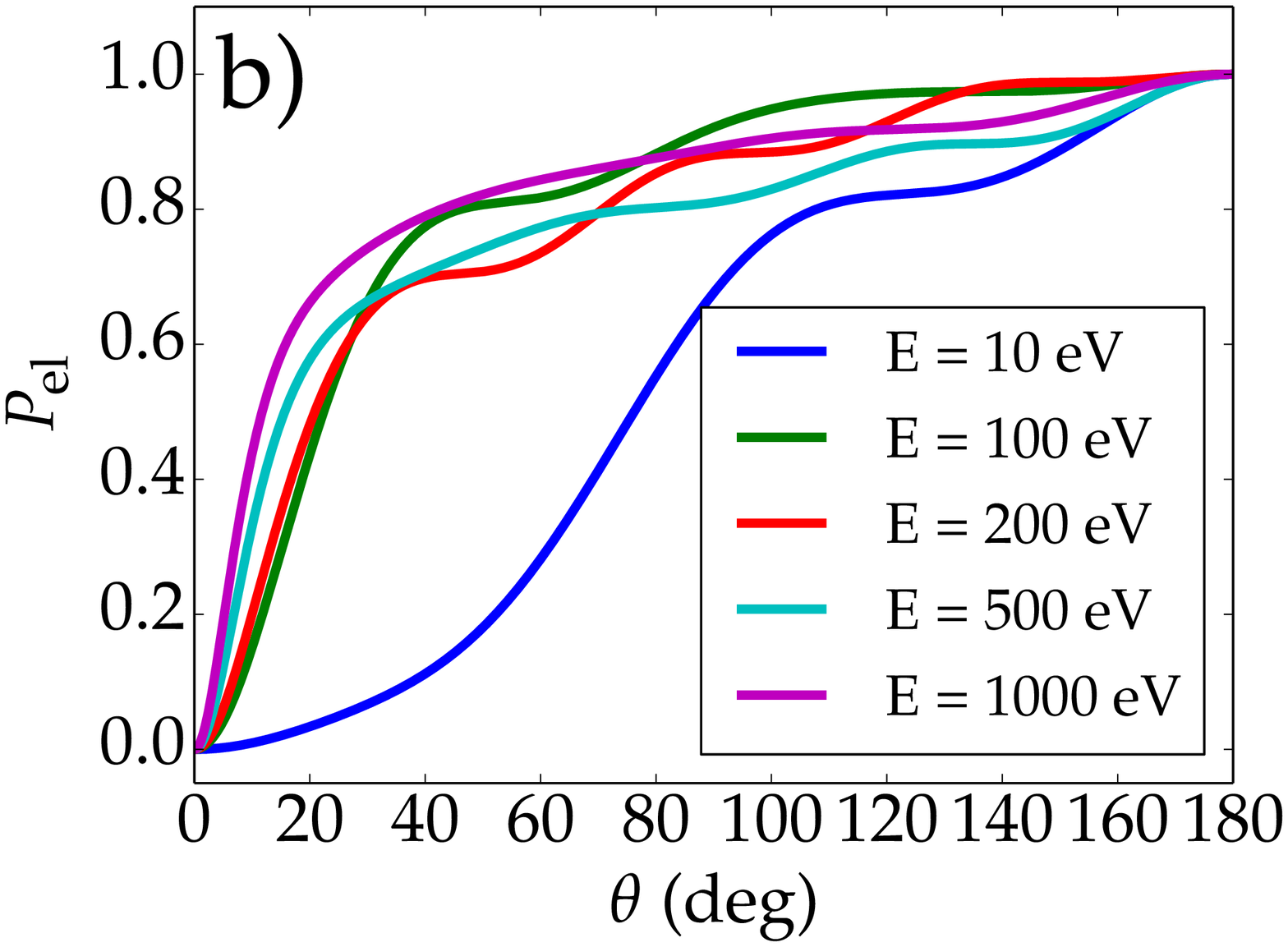}
    \includegraphics[width = 0.32\textwidth]{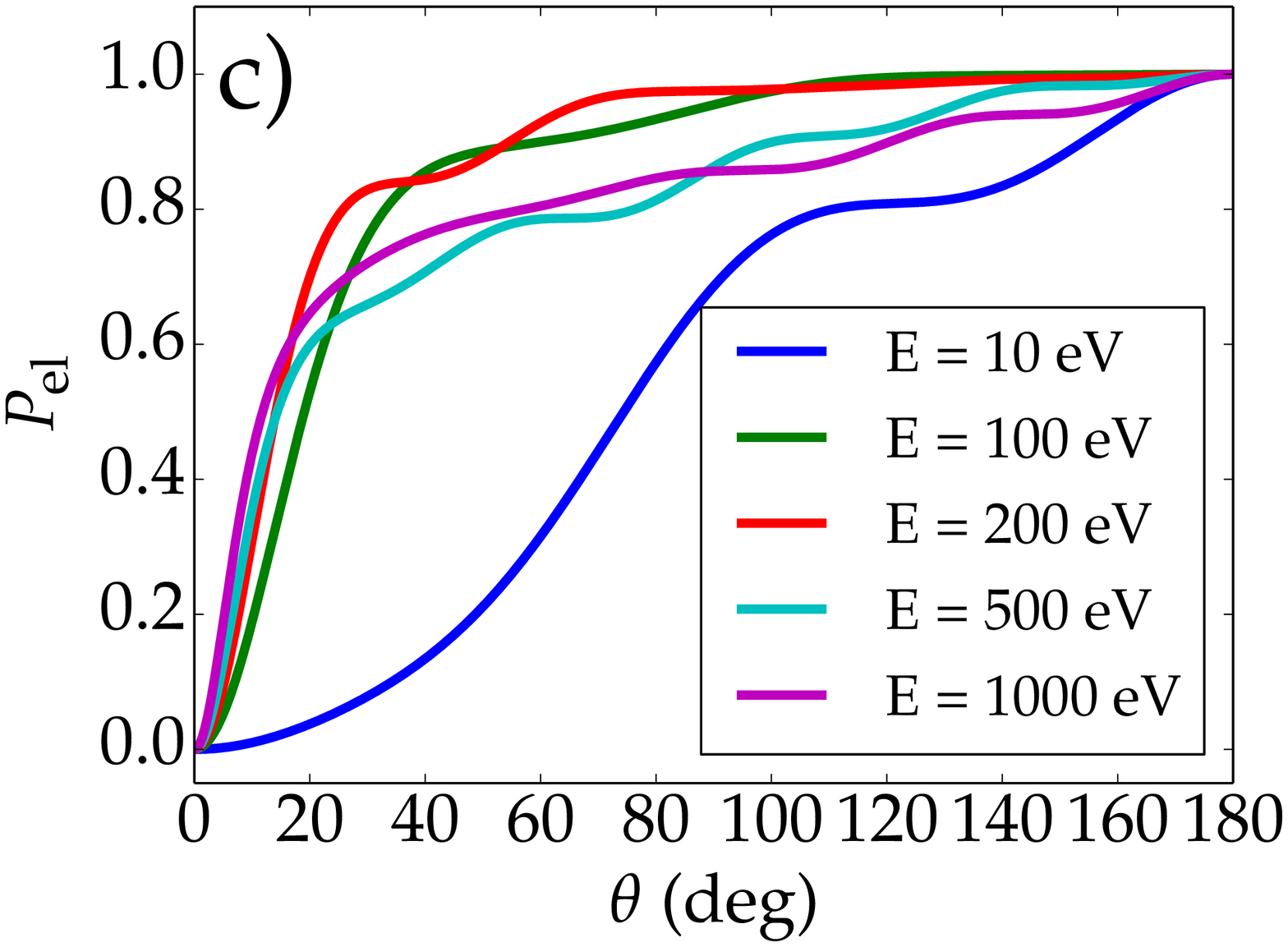}
    \caption{Cumulative elastic scattering probabilities of a) Cu, b) Ag, and c) Au as a function of the scattering angle for several kinetic energies.}
    \label{fig:PEL}
\end{figure}

From the knowledge of the function $\sigma_{el}(E)$, one can calculate the elastic scattering mean free path ($\lambda_\mathrm{el}$) at a given kinetic energy as follows: 
\begin{equation}
    \lambda_{\mathrm{el}}(E) = \frac{1}{N \sigma_\mathrm{el}(E)},
\end{equation}
where $N$ is the atomic density. 
In our MC model, the elastic scattering events lead to a change in the direction of the electron path. The scattering angle $\overline{\theta}$ after an elastic collision can be evaluated by calculating the cumulative elastic scattering probabilities $P_\mathrm{el}(\overline{\theta}, E)$ for different values of the electron kinetic energies $E$ (see Fig. \ref{fig:PEL}): 
\begin{equation}
    P_\mathrm{el}(\overline{\theta}, E) = \frac{2\pi}{\sigma_\mathrm{el}(E)}\int\limits_0^{\overline{\theta}} \frac{d\sigma_\mathrm{el}(E)}{d\Omega} \sin\theta d\theta
\end{equation}

Finally, $\overline{\theta}$ can be assessed by equalizing $P_\mathrm{el}$ at a given electron kinetic energy with a random number $\mu_3$, which is sampled uniformly in the interval [0,1]. 

\subsection{Inelastic scattering and secondary electron generation}
 
\begin{figure}[h!]
    \centering
    \includegraphics[width = 0.32\textwidth]{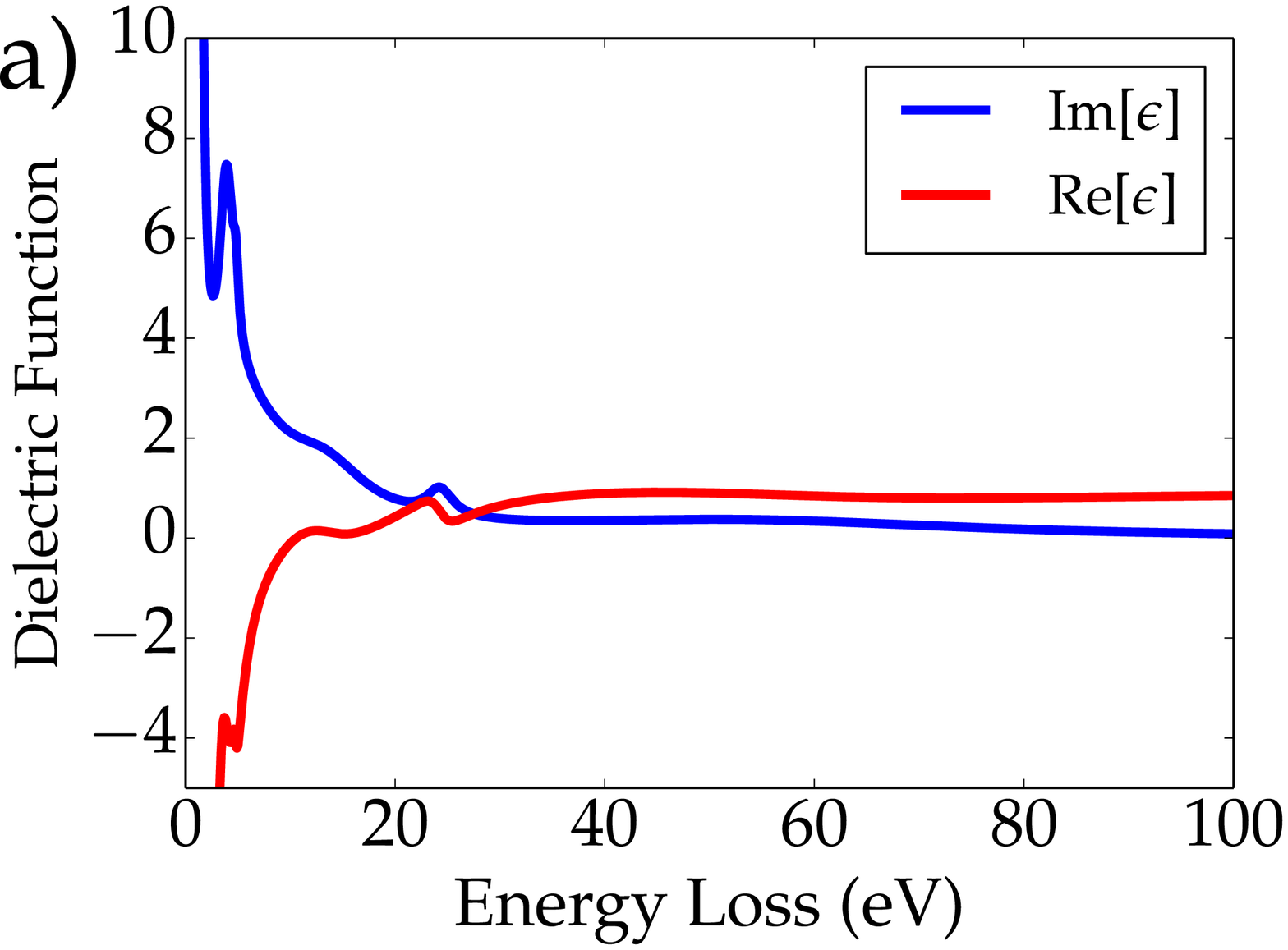}
    \includegraphics[width = 0.32\textwidth]{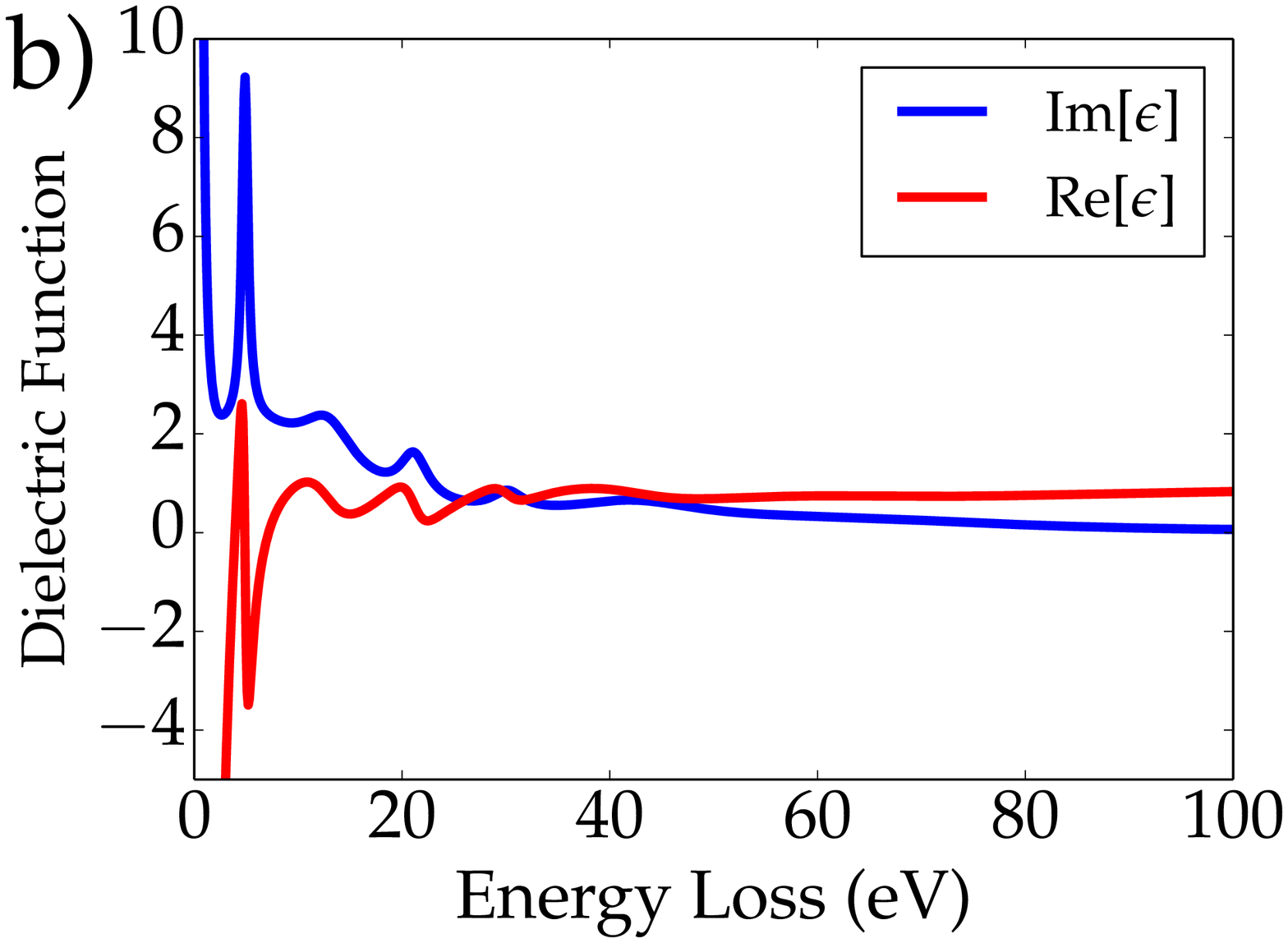}
    \includegraphics[width = 0.32\textwidth]{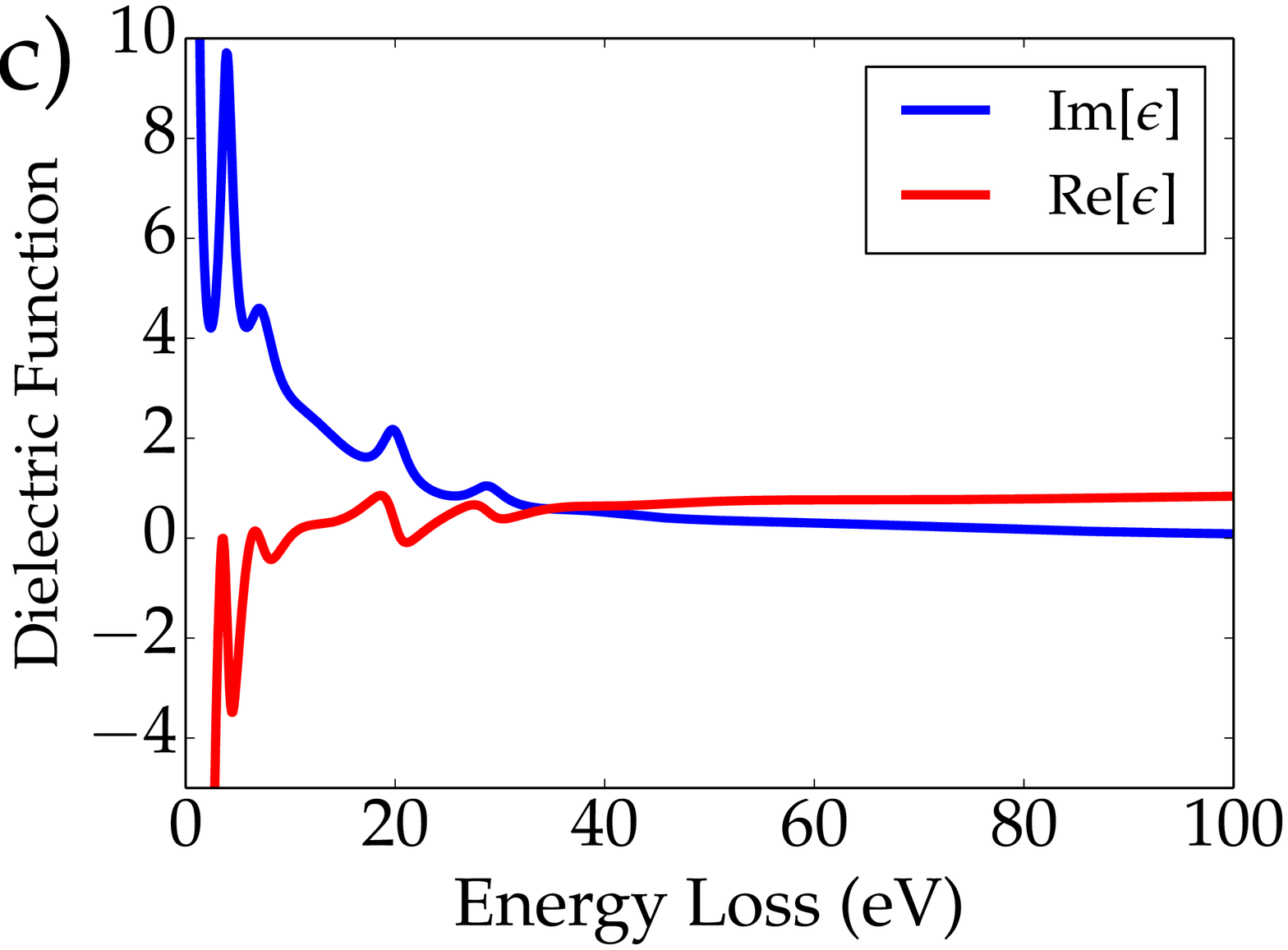}
    \caption{Real (red) and imaginary (blue) components of the dielectric functions of a) Cu a), b) Ag, and c) Au, as a function of energy loss for vanishing transferred momentum, obtained by using the fitting parameters reported in Ref. \citep{werner2009optical}. }
    \label{fig:epsilon}
\end{figure}

Inelastic scattering between the electrons in the beam and in the target atoms slow down the charge motion along the path.
The electrons moving within the solid may transfer a fraction of their kinetic energy to the target atomic electron cloud, producing both excitations and  ionizations. These processes can be fully described by the dielectric theory of Ritchie \citep{Ritchie_PhysRev_1957}. The dielectric function $\epsilon (W, \vec{q})$, which depends on transferred energy $W$ and momentum $\vec{q}$, describes the ``tendency'' of a solid to be polarized by an incoming charged particle or electromagnetic wave. The dielectric function can be assessed by both experiments and simulations \citep{AZZOLINI2017299,doi:10.1002/pssb.200982339}. In this case, we decided to use the dielectric function obtained from experimental reflection electron energy loss spectra by Werner {\it et al.} \citep{werner2009optical}, as it includes both the bulk and the surface contributions to the electronic excitation. The real and imaginary components were fitted via Drude-Lorentz functions, which mimic plasmon oscillations, whose fitting parameters are provided in Ref. \citep{werner2009optical}. The resulting real and imaginary components of the dielectric functions of Cu, Ag and Au for vanishing transferred momentum are shown in Fig.\ref{fig:epsilon}. 

\begin{figure}[h!]
    \centering
    \includegraphics[width = 0.32\textwidth]{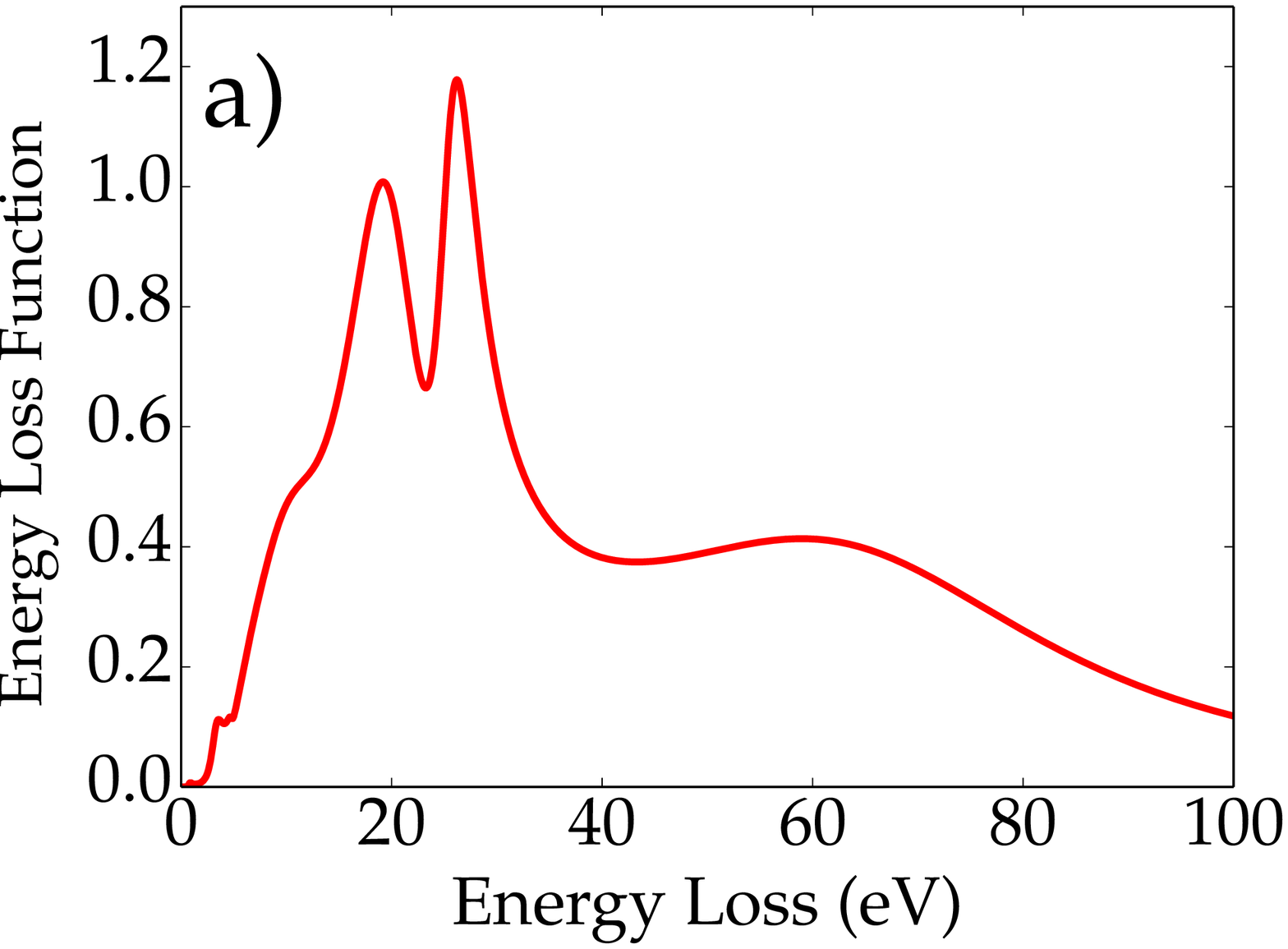}
    \includegraphics[width = 0.32\textwidth]{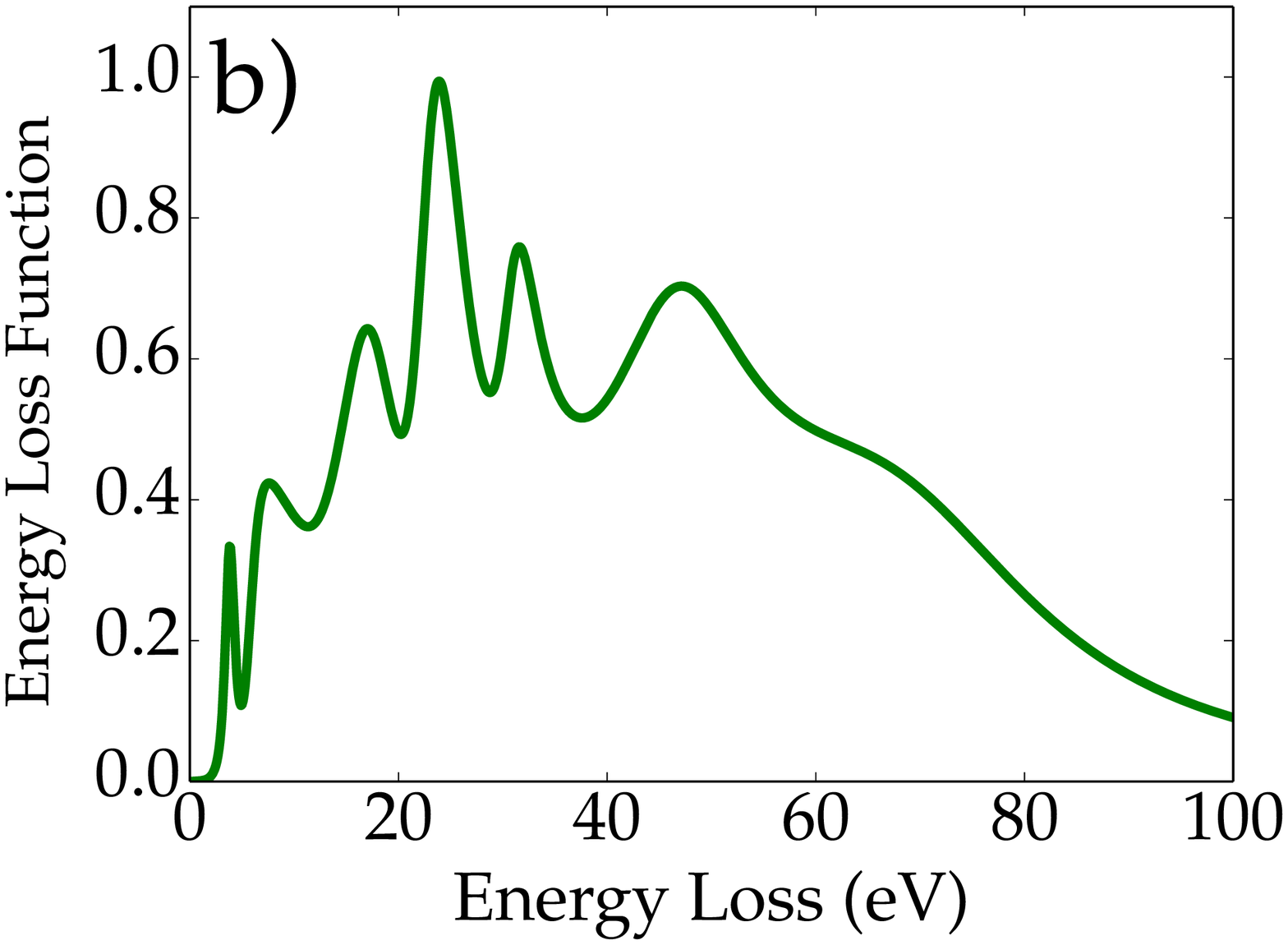}
    \includegraphics[width = 0.32\textwidth]{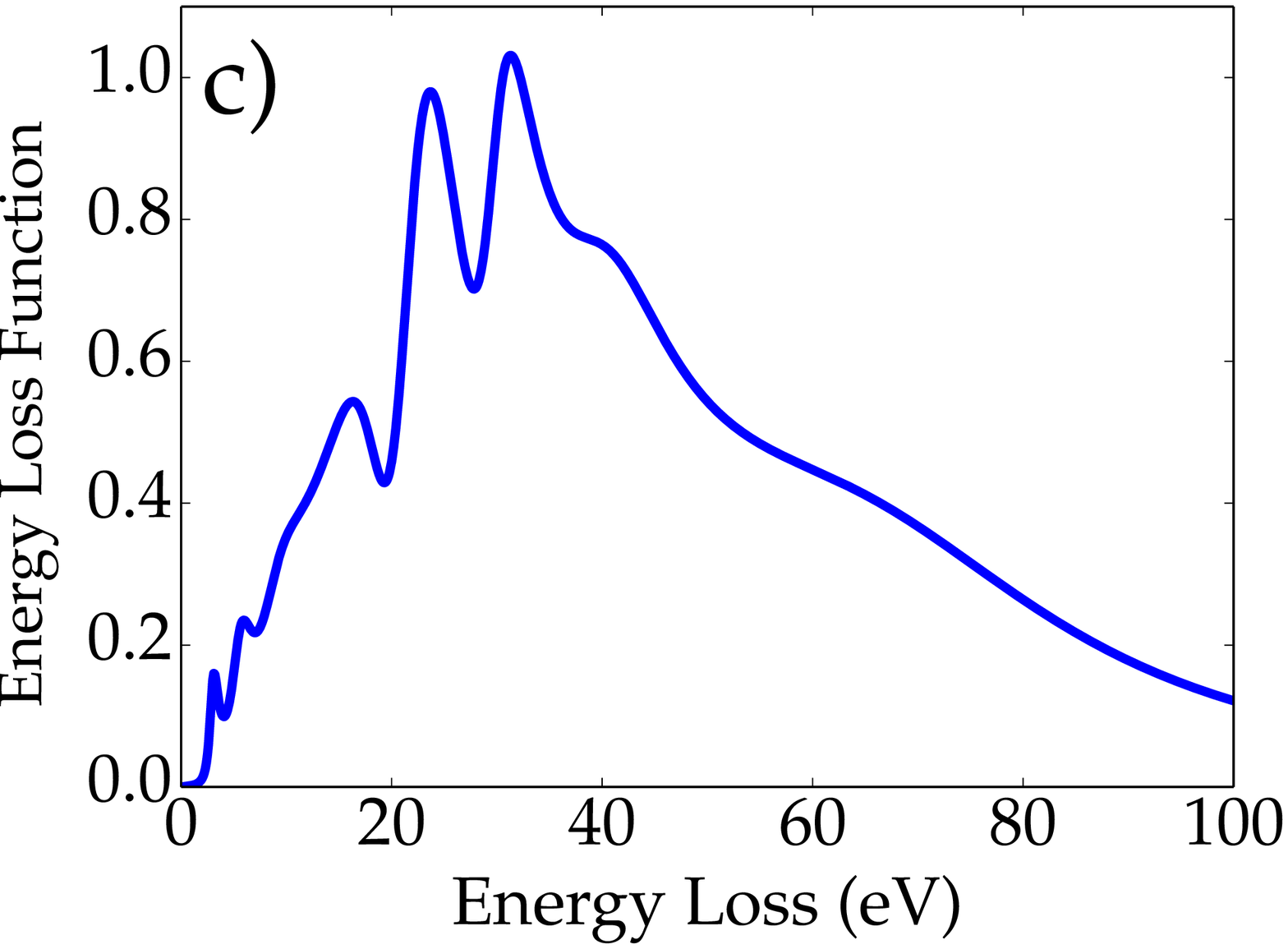}
    \caption{Energy Loss Functions of Cu a), Ag b) and Au c), as a function of energy loss and for vanishing transferred momentum obtained using the fitting parameters provided in Ref. \citep{werner2009optical}.}
    \label{fig:ELF}
\end{figure}

From the knowledge of the dielectric function, one can obtain the key quantity in charge transport Monte Carlo simulations, that is the Energy Loss Function (ELF) defined by the following relation:
\begin{equation}\label{ELF}
    ELF = \mathrm{Im}\left[-\frac{1}{\epsilon(q, W)}\right]= \frac{\mathrm{Im[}\epsilon]}{\mathrm{Re[}\epsilon]^2+\mathrm{Im[}\epsilon]^2}
\end{equation}
Using the formula \ref{ELF}, the ELF of Cu, Ag, and Au were calculated and are shown in Fig. \ref{fig:ELF}.

\begin{figure}[h!]
    \centering
    \includegraphics[width = 0.32\textwidth]{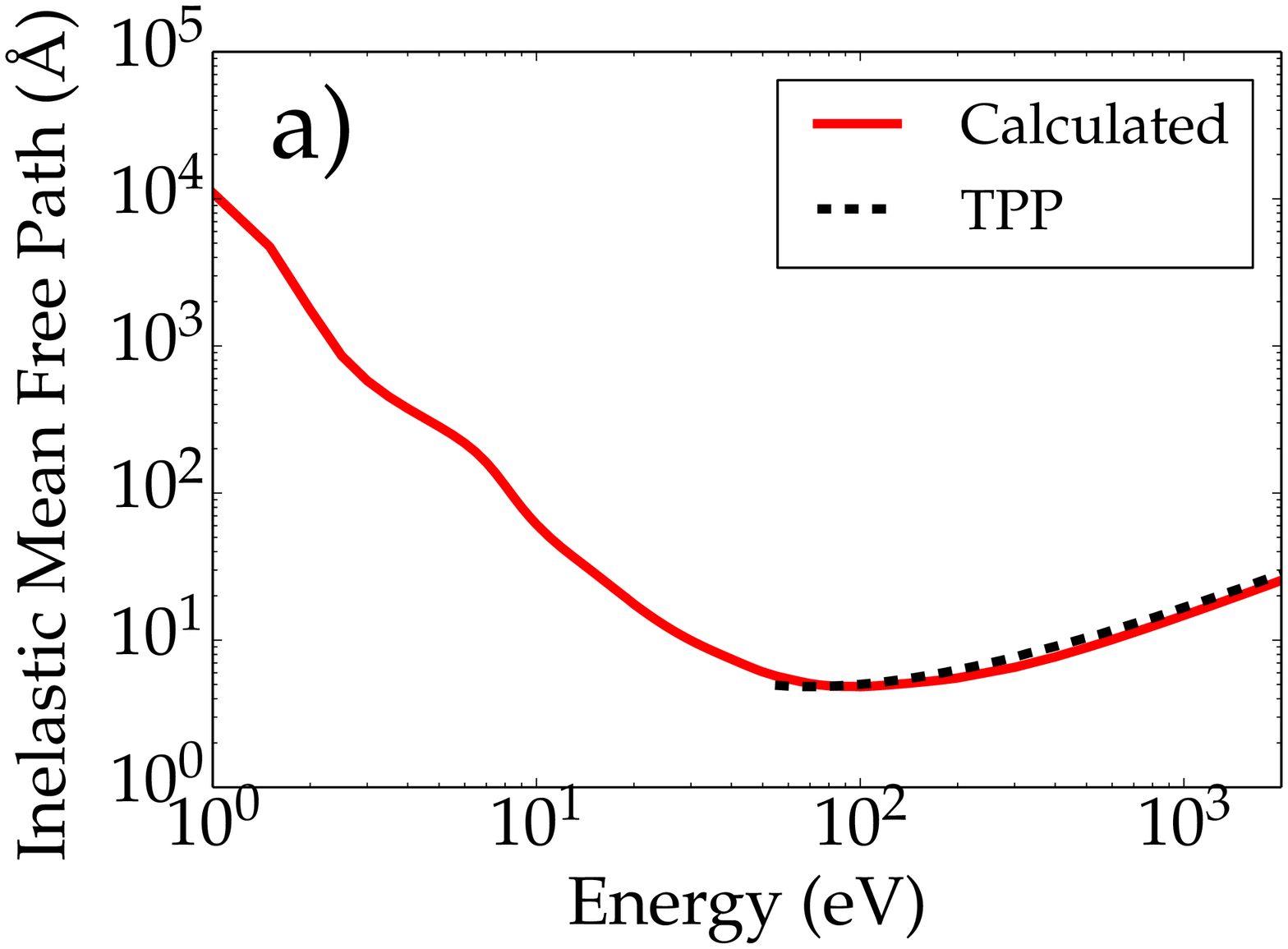}
    \includegraphics[width = 0.32\textwidth]{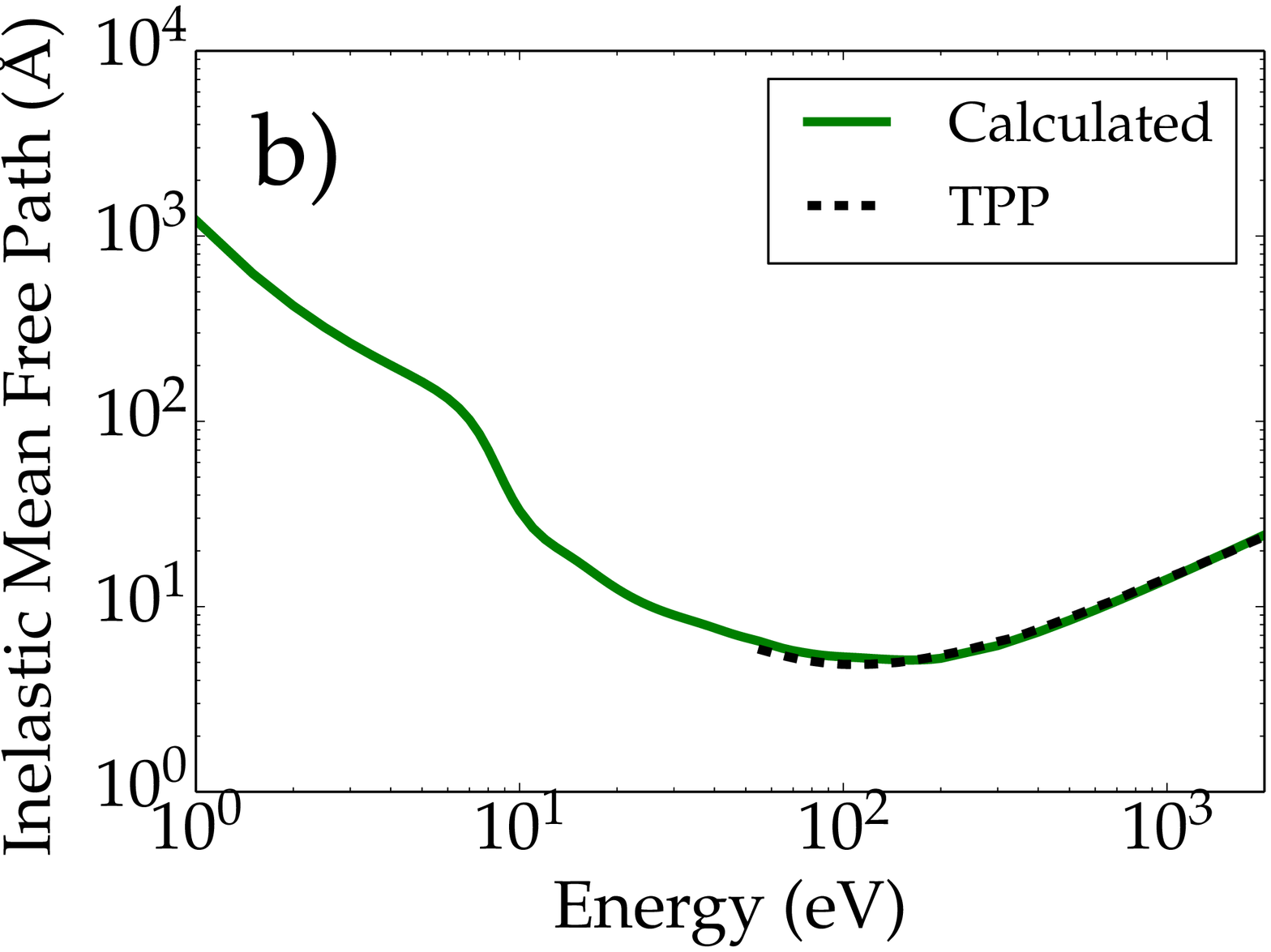}
    \includegraphics[width = 0.32\textwidth]{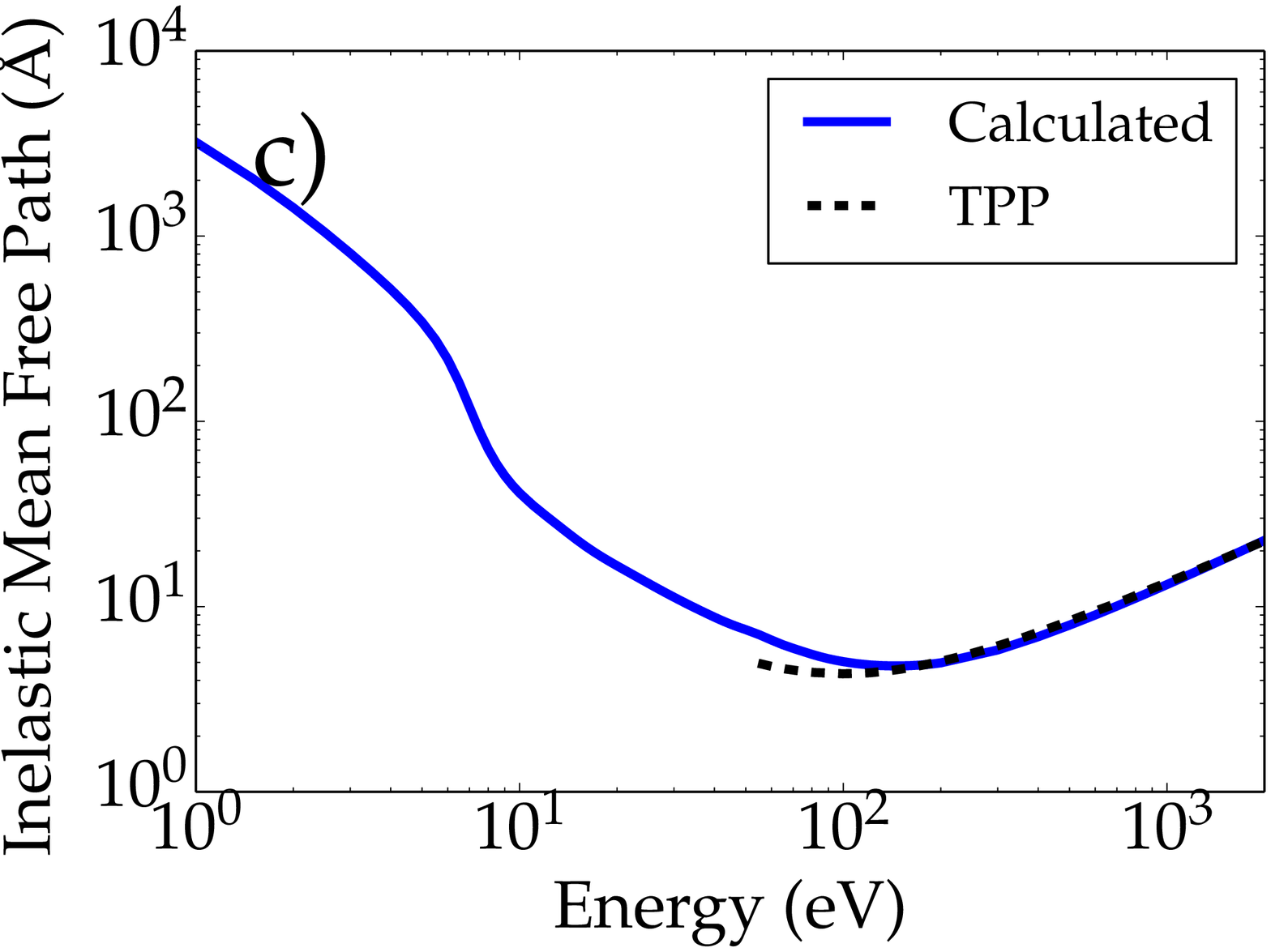}
    \caption{Inelastic Mean Free Paths of a) Cu, b) Ag, and c) Au as a function of the electron kinetic energy. Dashed lines show the data obtained by Tanuma et al. \citep{Tanuma_SIA_2009}.}
    \label{fig:imfp}
\end{figure}

To extend the dielectric function to  transferred momenta $\vec{q}$ different from zero we apply the following dispersion law to the characteristic energies of the oscillators $W_i(q)$:
\begin{equation}
    W_i(q) = W_i(q = 0) + \alpha\frac{(\hbar q^2)}{2m }
\end{equation}
where $m$ is the electron mass, $q$ the modulus of the transferred momentum, and $\alpha$ a dispersion coefficient that depends on the energy scale. 
Indeed, the Drude-Lorentz theory was developed to describe the excitation in the low energy region, corresponding to energy losses lower than the semi-core transition energies. To extend properly the Drude-Lorentz theory to higher energies the $\alpha$ dispersion parameter must be tuned. 
According to Ref. \citep{werner2009optical}, we set $\alpha=1$ for oscillator energies $W_i(q = 0)$ lower than the characteristic energies of the semi-core transitions. For larger oscillators energy, $\alpha$ was set to 0.5, a valiue which ensures the best agreement with experimental value \citep{werner2009optical}.
The threshold energies of the semi-core transitions are $E_{3p3/2} = \ $ 75.1 eV for Cu, $E_{5p3/2} = \ $  57.2 eV for Ag, and $E_{4p3/2} = \ $58.3 eV for Au  respectively \citep{werner2009optical}. \\
\indent The knowledge of the ELF can be used for calculating the differential inverse inelastic mean free path (DIIMFP), which is defined as: 
\begin{equation}
    \frac{d \lambda_\mathrm{inel}^{-1}}{dW} = \frac{1}{\pi a_0 E}\int_{q_-}^{q_+} \frac{dq}{q} \mathrm{Im}\left[-\frac{1}{\epsilon(q, W)}\right]
\end{equation}
where $a_0$ is the Bohr radius and the integration limits are $q_\pm = \sqrt{2mE} \pm \sqrt{2m(E-W)}$, with $W$ the transferred energy. 
The total inelastic mean free path $\lambda_{\mathrm{inel}}$ (IMFP) is obtained by integrating the DIIMFP in the energy loss range $[0, E/2]$. In Figs. \ref{fig:imfp}a), b), c) we present the IMFPs of Cu, Ag, and Au respectively compared with the simulations performed by Tanuma et al. \citep{Tanuma_SIA_2009}.

\begin{figure}[h!]
    \centering
    \includegraphics[width = 0.32\textwidth]{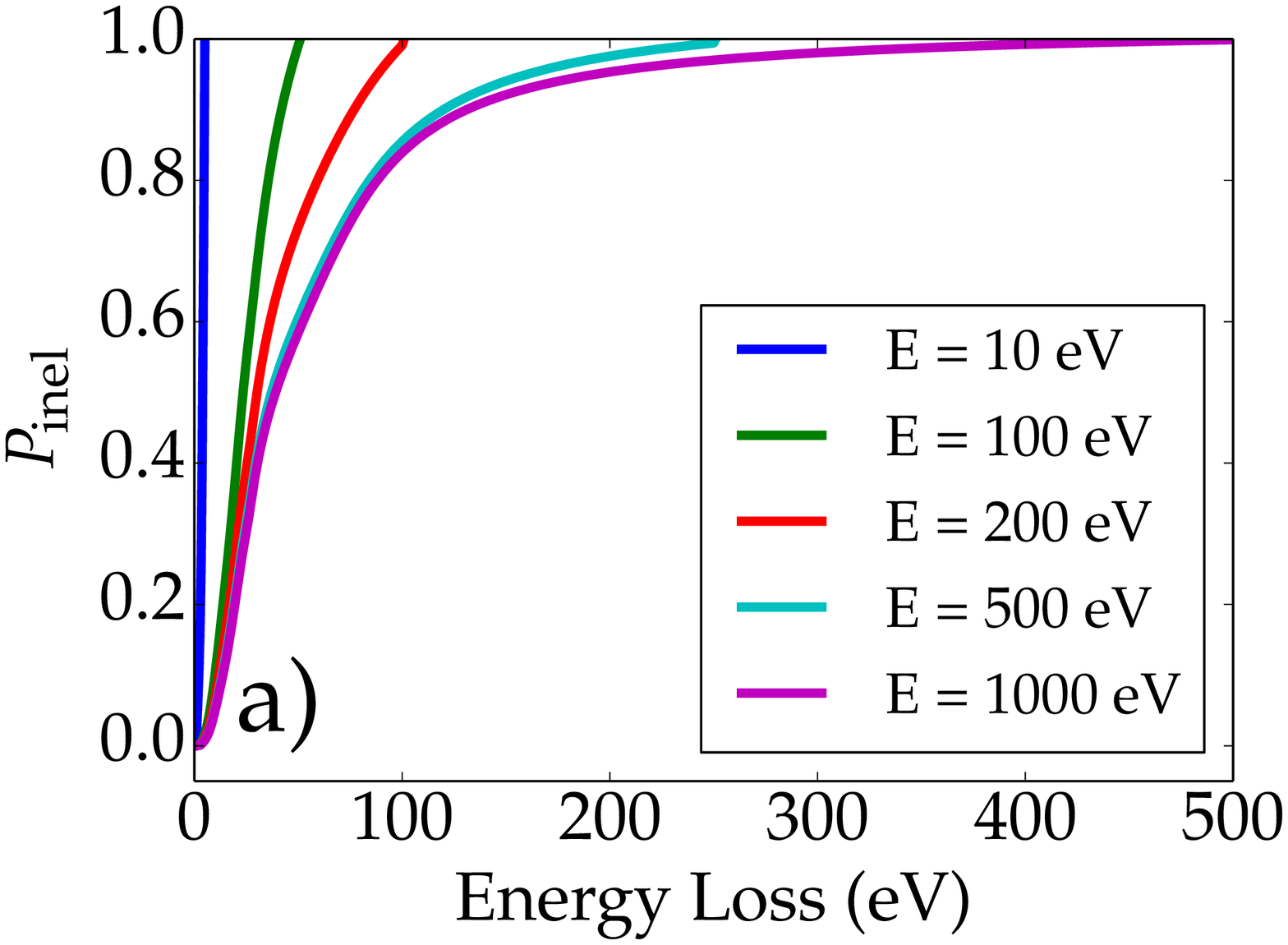}
    \includegraphics[width = 0.32\textwidth]{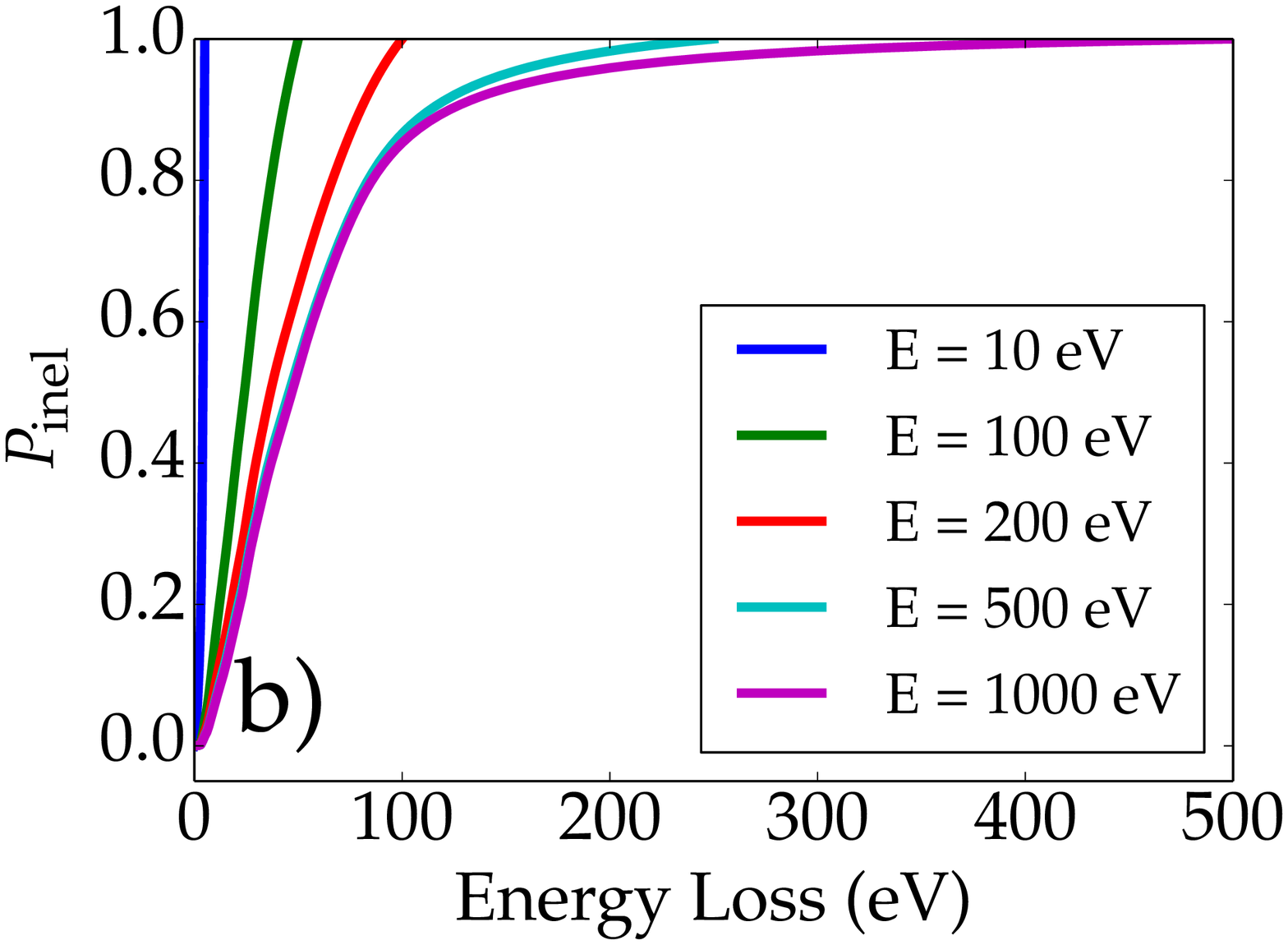}
    \includegraphics[width = 0.32\textwidth]{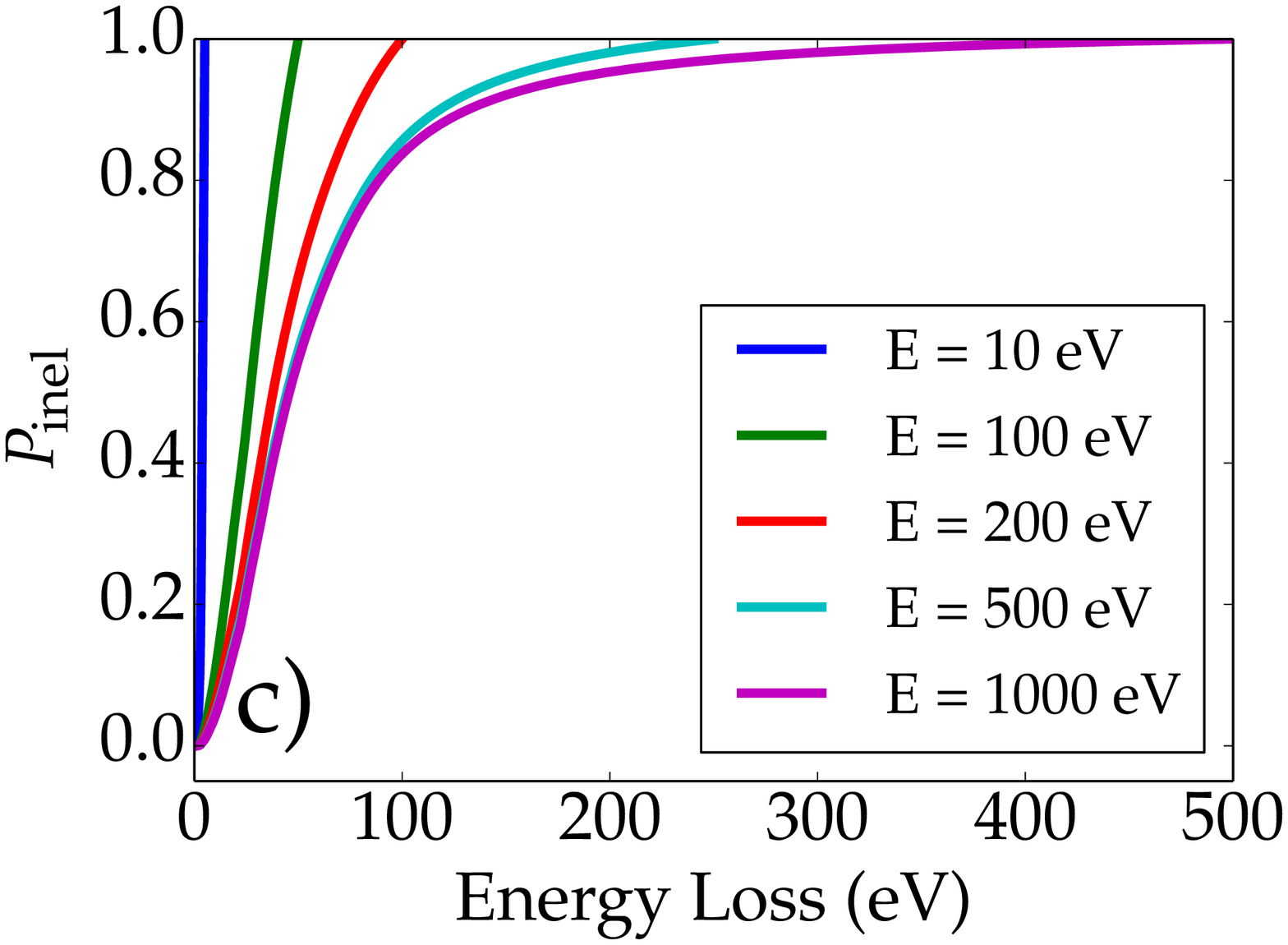}
    \caption{Cumulative inelastic scattering probabilities of a) Cu, b) Ag, and c) Au as a function of the energy loss for different kinetic energies}
    \label{fig:pinel}
\end{figure}

The tendency of electrons to undergo elastic or inelastic collisions can be assessed by their respective distribution probabilities, $p_\mathrm{el}={\lambda_\mathrm{tot}}/{\lambda_{\mathrm{el}}}$ and $p_\mathrm{inel} = {\lambda_\mathrm{tot}}/{\lambda_{\mathrm{inel}}}$, while the total mean free path $\lambda$ is defined by the following relation:
\begin{equation}
    \lambda^{-1}(E) = \lambda_\mathrm{inel}^{-1}(E) + \lambda_\mathrm{el}^{-1}(E)
\end{equation}
The MC scheme applied to the electron transport within solids proceeds in the following way: a random number $\mu_4$ uniformly distributed in the range $[0,1]$ is generated and compared with $p_\mathrm{inel}$. If the condition $r < p_\mathrm{inel}$ is satisfied the collision is classified as inelastic, otherwise is elastic.\\ 
\indent As a results of an inelastic scattering, the impinging electron loses a fraction $\overline{W}$ of its kinetic energy. In the MC calculation, the value of $\overline{W}$ is determined for each inelastic collision by comparing a random number $\mu_5$ (uniformly distributed in the interval $[0,1]$), with the correspondent cumulative probability distribution $P_\mathrm{inel}(E,W)$.
These probabilities are calculated for different values of the electron kinetic energies $E$ (see Fig. \ref{fig:pinel}), as: 
\begin{equation}
    P_\mathrm{inel}(E,\overline{W}) = \lambda_\mathrm{inel}(E) \int_0^{\overline{W}} \frac{d \lambda_\mathrm{inel}^{-1}(E,W)}{dW} dW
\end{equation}
The value of $\overline{W}$ for which the value of $  P_\mathrm{inel}$ is equal to $\mu_5$ is the energy loss upon one inelastic collision. Thus, the electron kinetic energy will be decreased by this value. The angular deviation due to inelastic scattering is evaluated according to the classical binary collision theory.\\
\indent Moreover, should the energy loss be larger than the first ionization energy $B$ (that is, the energy required to extract one electron from the outern electron shell of the target atom), a secondary electron is emitted with kinetic energy equal to $\bar{W}-B$. After the ionization event, the generated secondary electron moves inside the solid target as any other particle, due to indistinguishability of electrons.
\begin{figure}[h!]
    \centering
    \includegraphics[width = 0.32\textwidth]{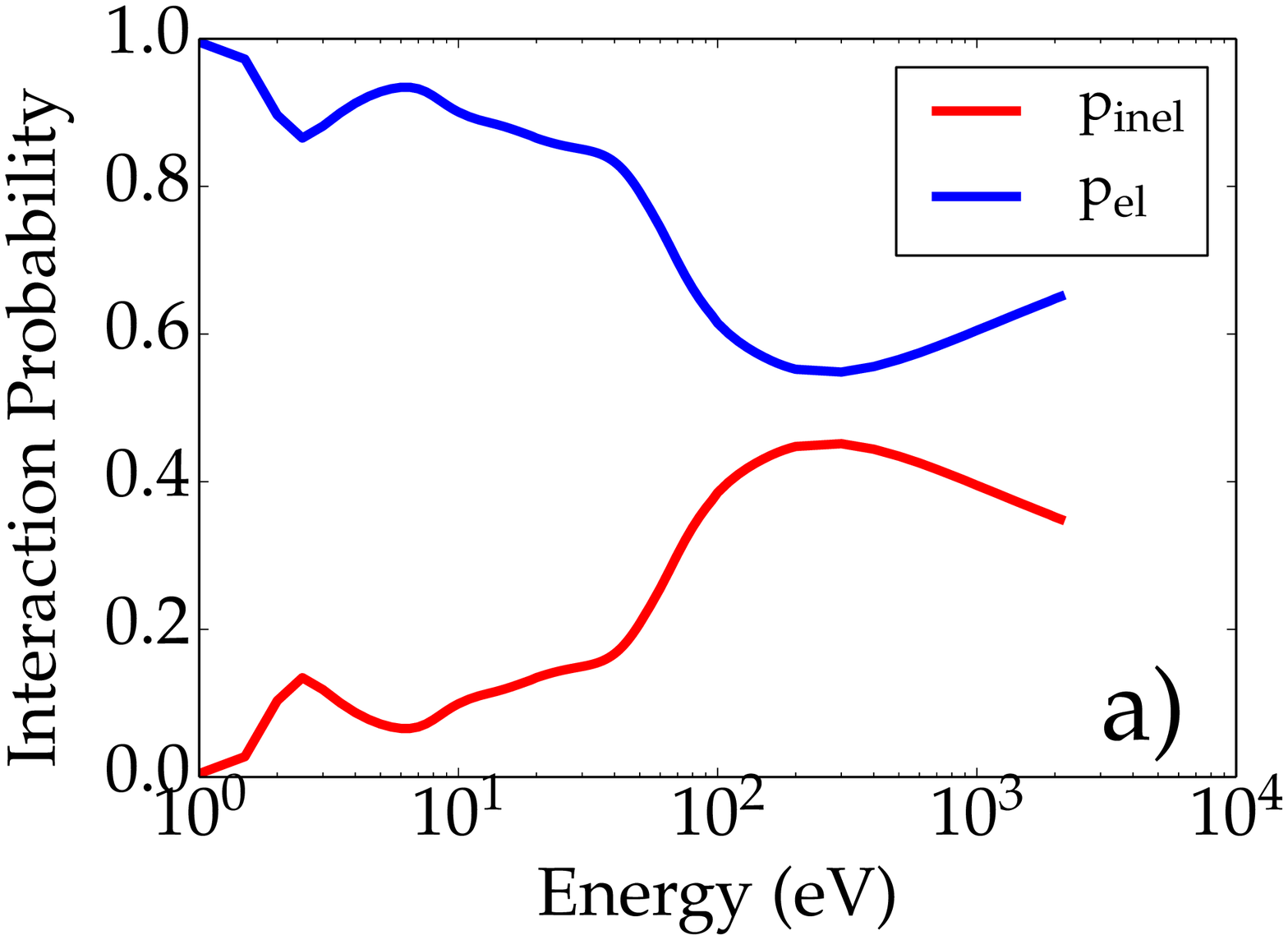}
    \includegraphics[width = 0.32\textwidth]{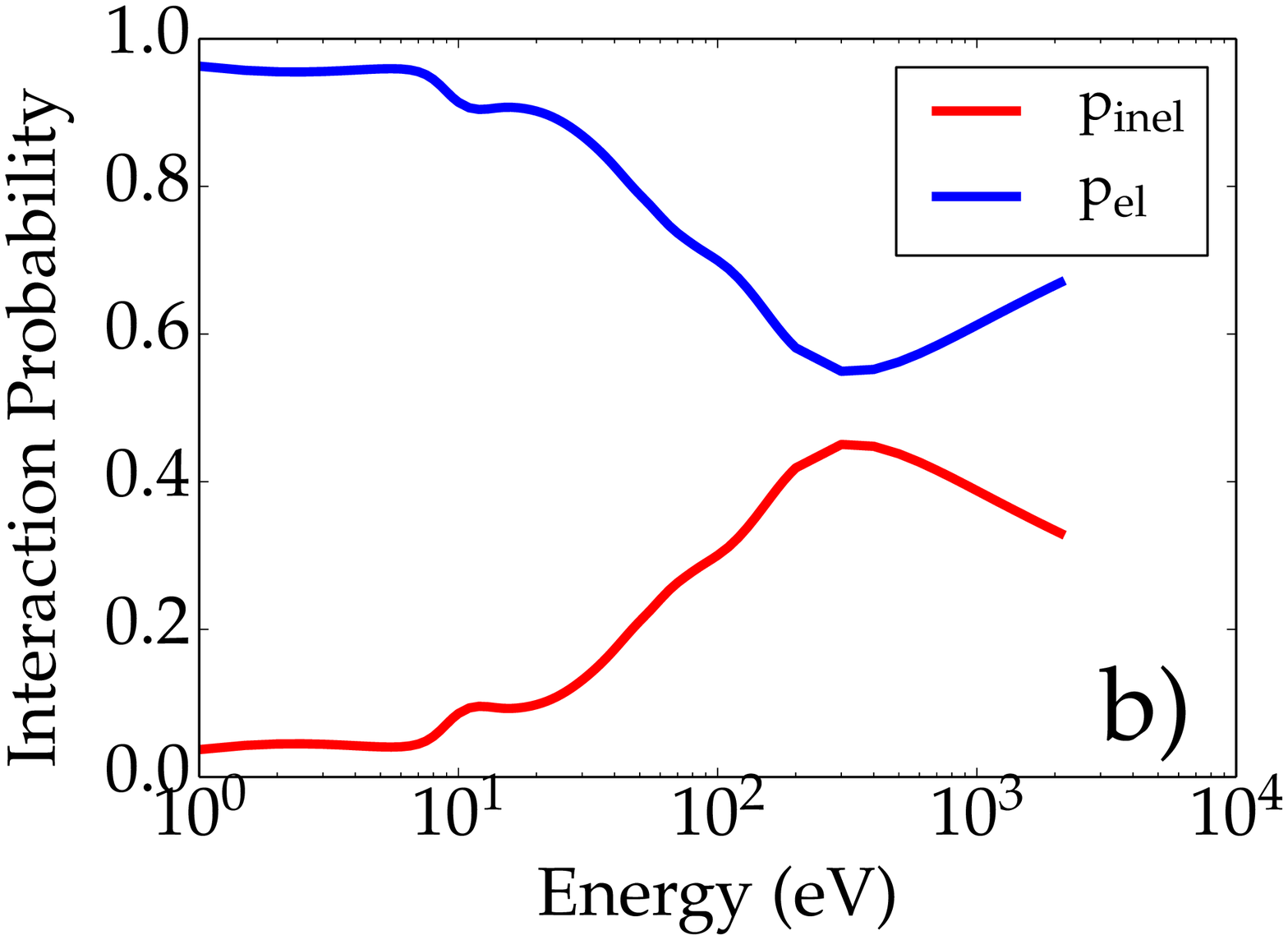}
    \includegraphics[width = 0.32\textwidth]{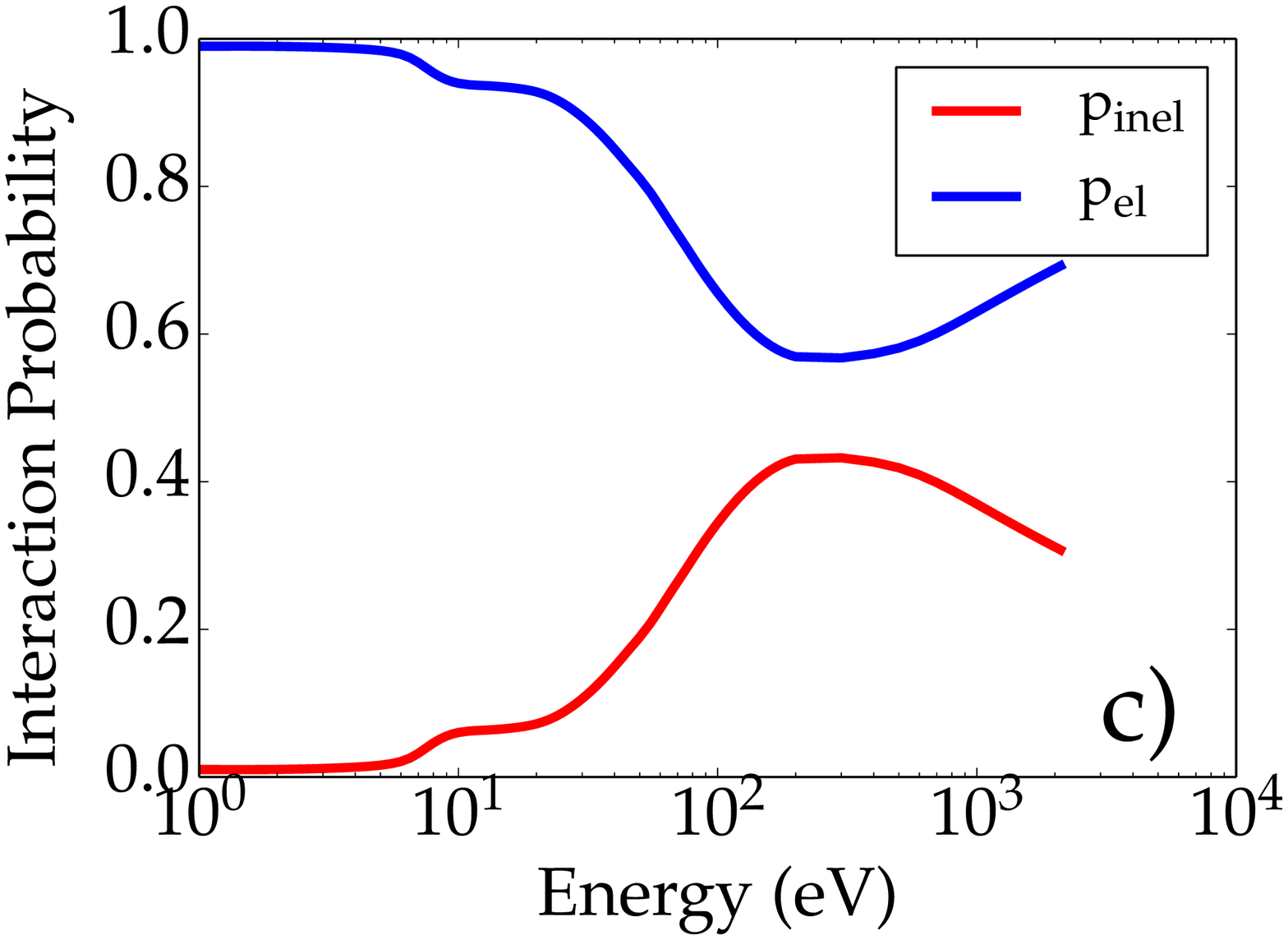}
    \caption{Scattering probabilities for of a) Cu a), b) Ag, and c) Au as a function of the electron energy. }
    \label{fig:prob}
\end{figure}

\section{Results and discussion}

In our MC simulations we considered copper, silver and gold bulk metals as test cases of our method. In Tab. \ref{tab:char} the characteristics of these materials are reported. In the next sections we present both electron emission spectra and yield curves of these three metals. 
\begin{table}[h!]
    \centering
   \begin{tabular}{c|c c c}
     Metal &  density (g/cm$^3$) &  $<B>$ (eV)\\  \hline
     Cu & 8.96 \citep{montanari2007calculation} &  7.726\\
     Ag & 10.5\citep{Tanuma_SIA_2009}  &  7.576\\
     Au & 19.32\citep{denton2008influence} &   9.226\\

\end{tabular}
    \caption{Characteristic quantities of target materials: the target density and the mean ionization energy characteristic of each sample.}
    \label{tab:char}
\end{table}

\subsection{Full energy emission spectra of Cu}

MC simulations were performed to calculate the full energy emission spectrum of bulk copper for different energies $E$ of the primary beam. The number of electrons in the beam was set to 10$^7$ to obtain stable results. 
The sample work function was set to 5.4 eV (as will be explained below, the experimental value is equal to 4.6 eV).\\ 
\indent The initial electron kinetic energy is distributed as by the reference experimental elastic peak of copper reported in the previous Fig. 1. Emitted electrons
are collected as a function of their kinetic energies. Theoretical spectra are compared with our experimental data for 
different initial kinetic energies in Fig. 9. 
The panels on the left side of the figure report spectra normalized at a common height of the secondary  
electron (SE) emission peak, while the right panels show the spectra normalized at a common area. 
\begin{figure*}[p]
    \centering
    \includegraphics[width = 0.40\textwidth]{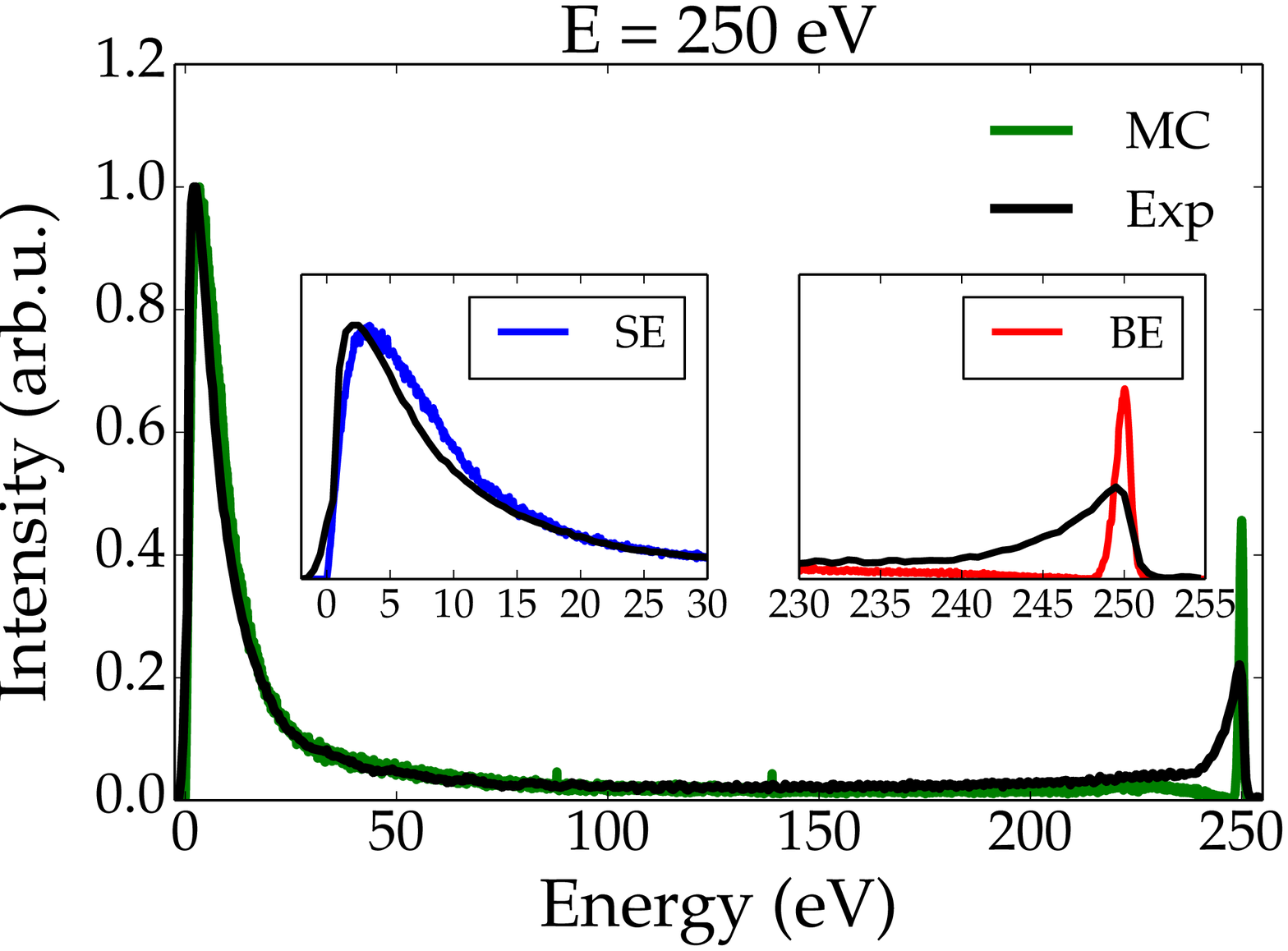}
    \includegraphics[width = 0.40\textwidth]{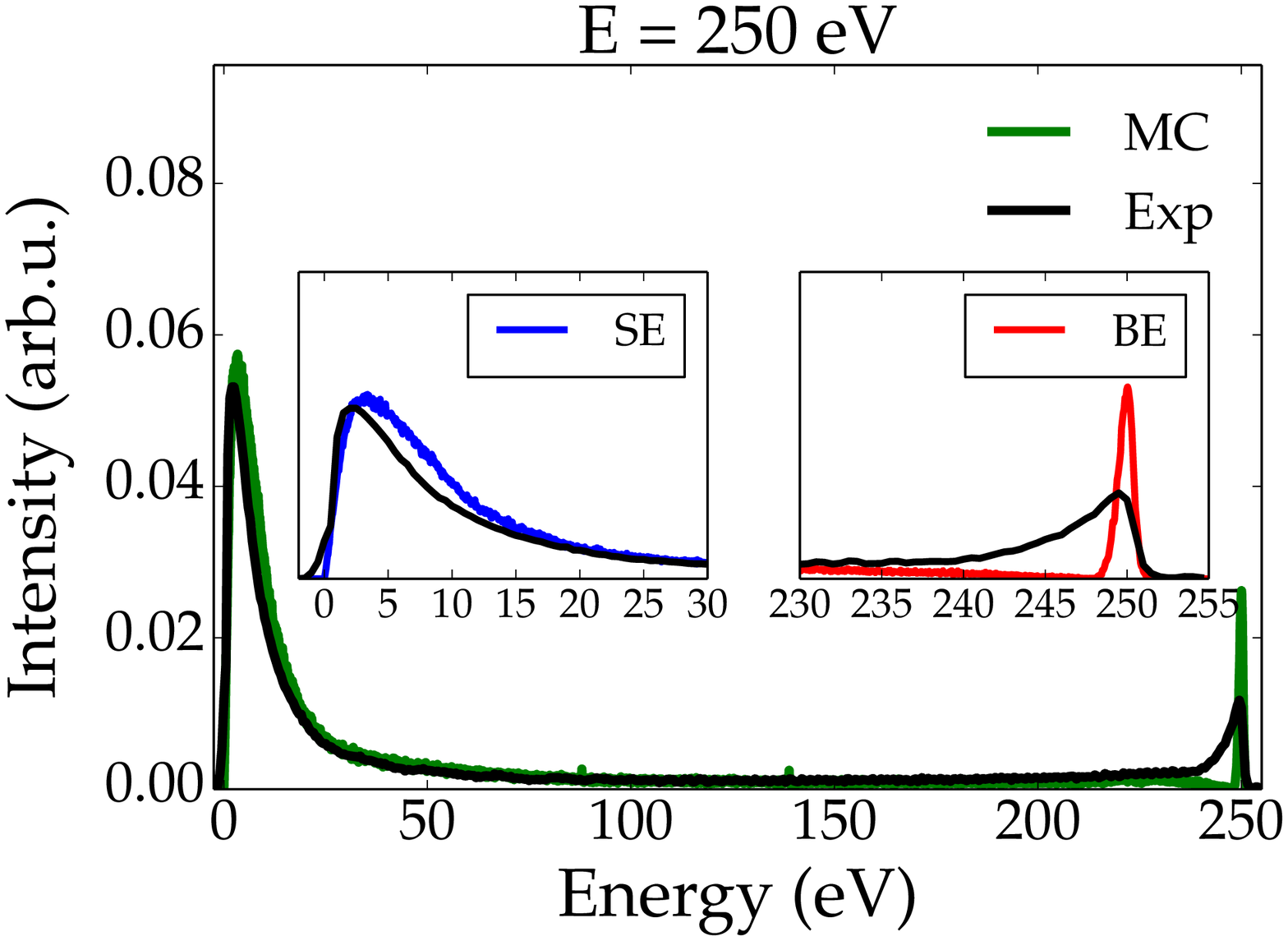}\\
    \includegraphics[width = 0.40\textwidth]{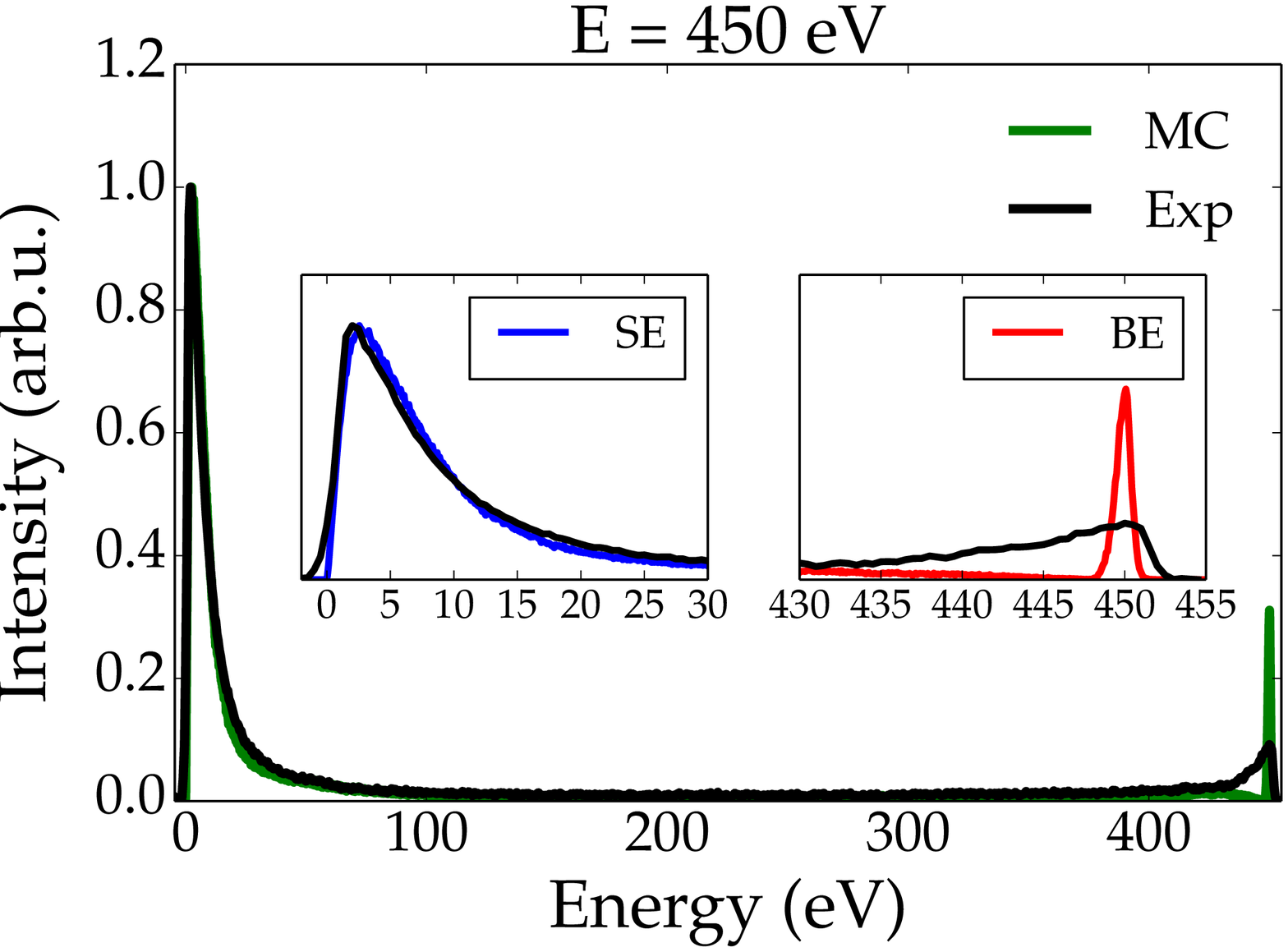}
    \includegraphics[width = 0.40\textwidth]{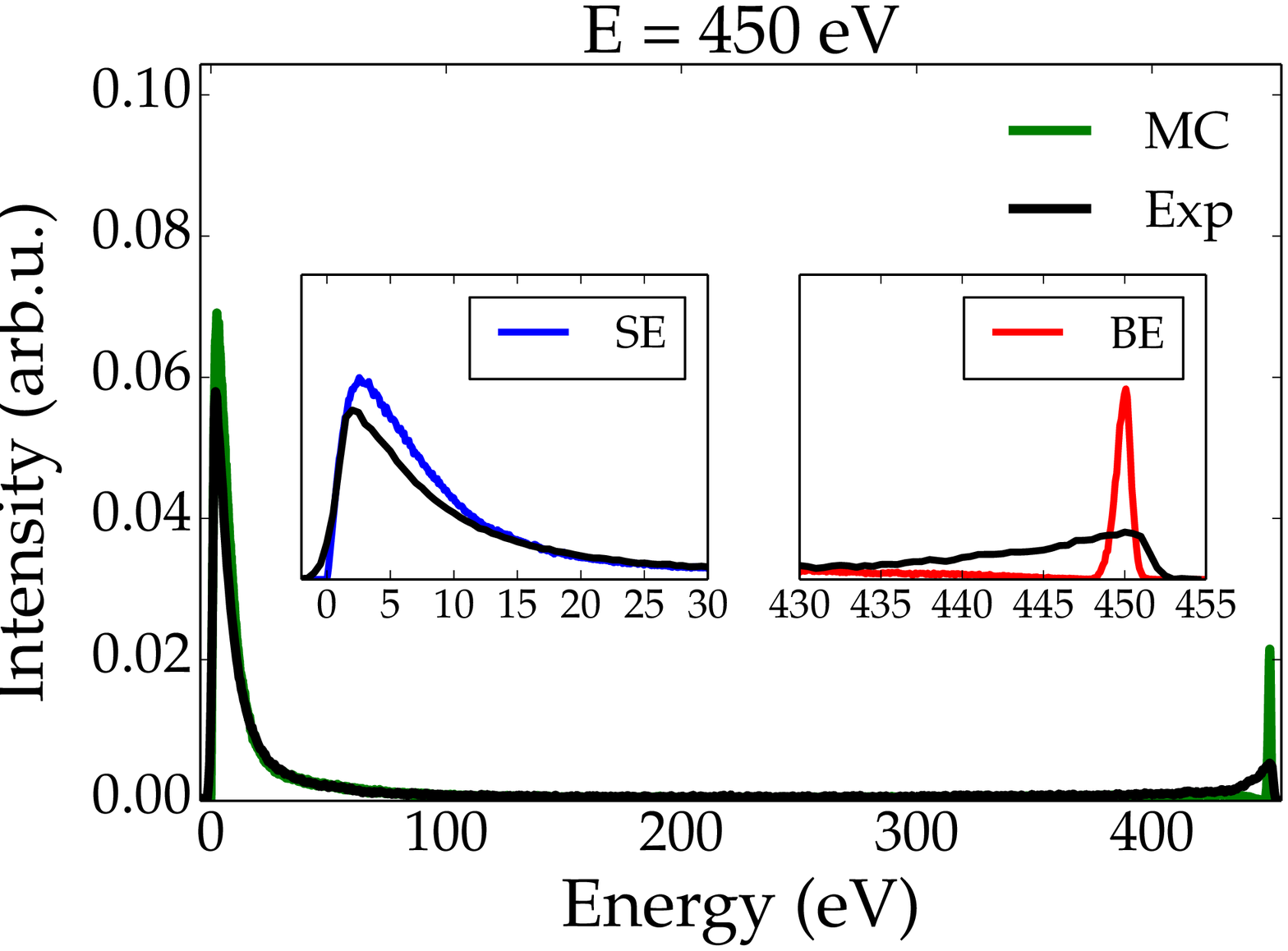}\\
     \includegraphics[width = 0.40\textwidth]{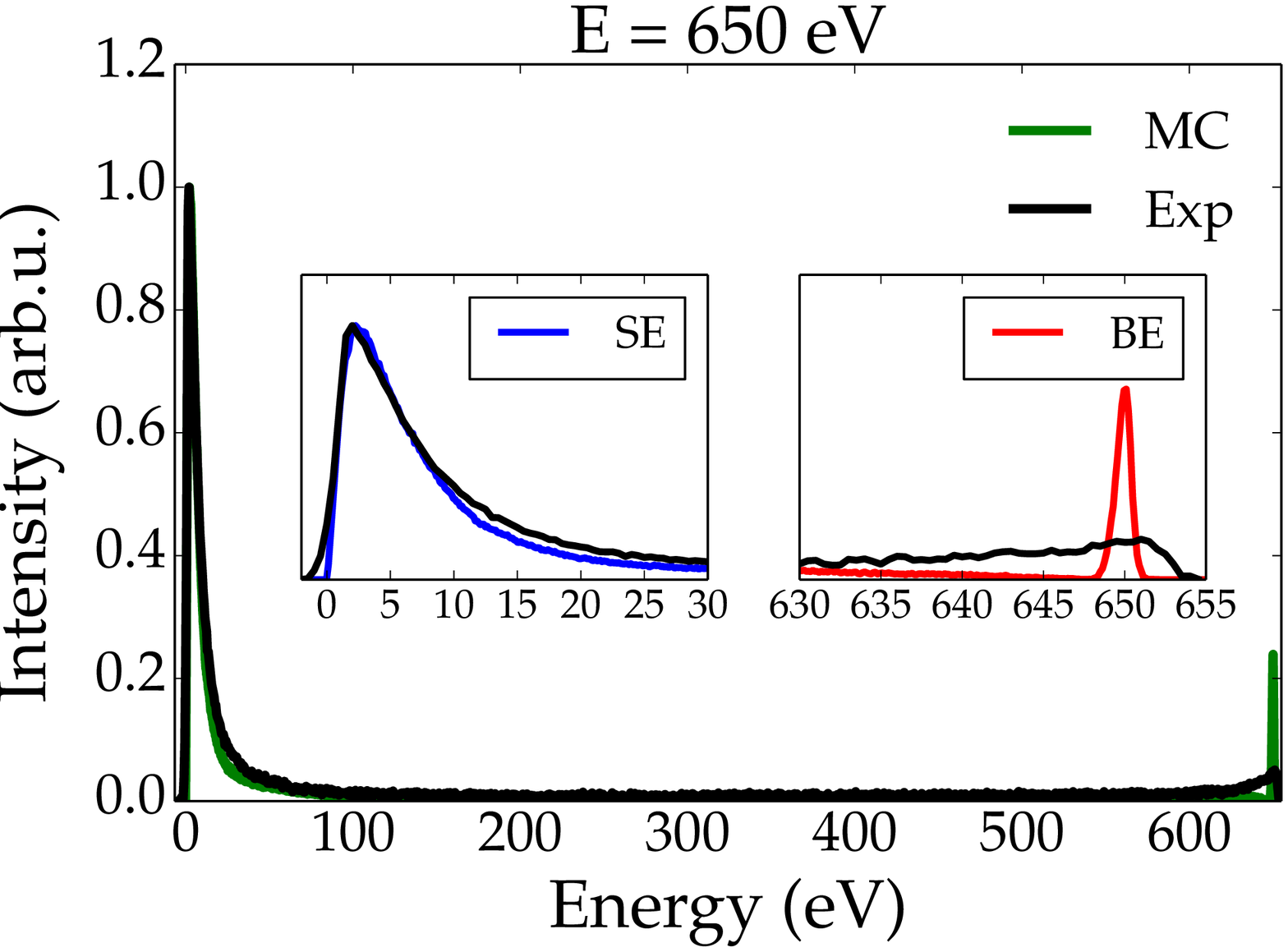}
    \includegraphics[width = 0.40\textwidth]{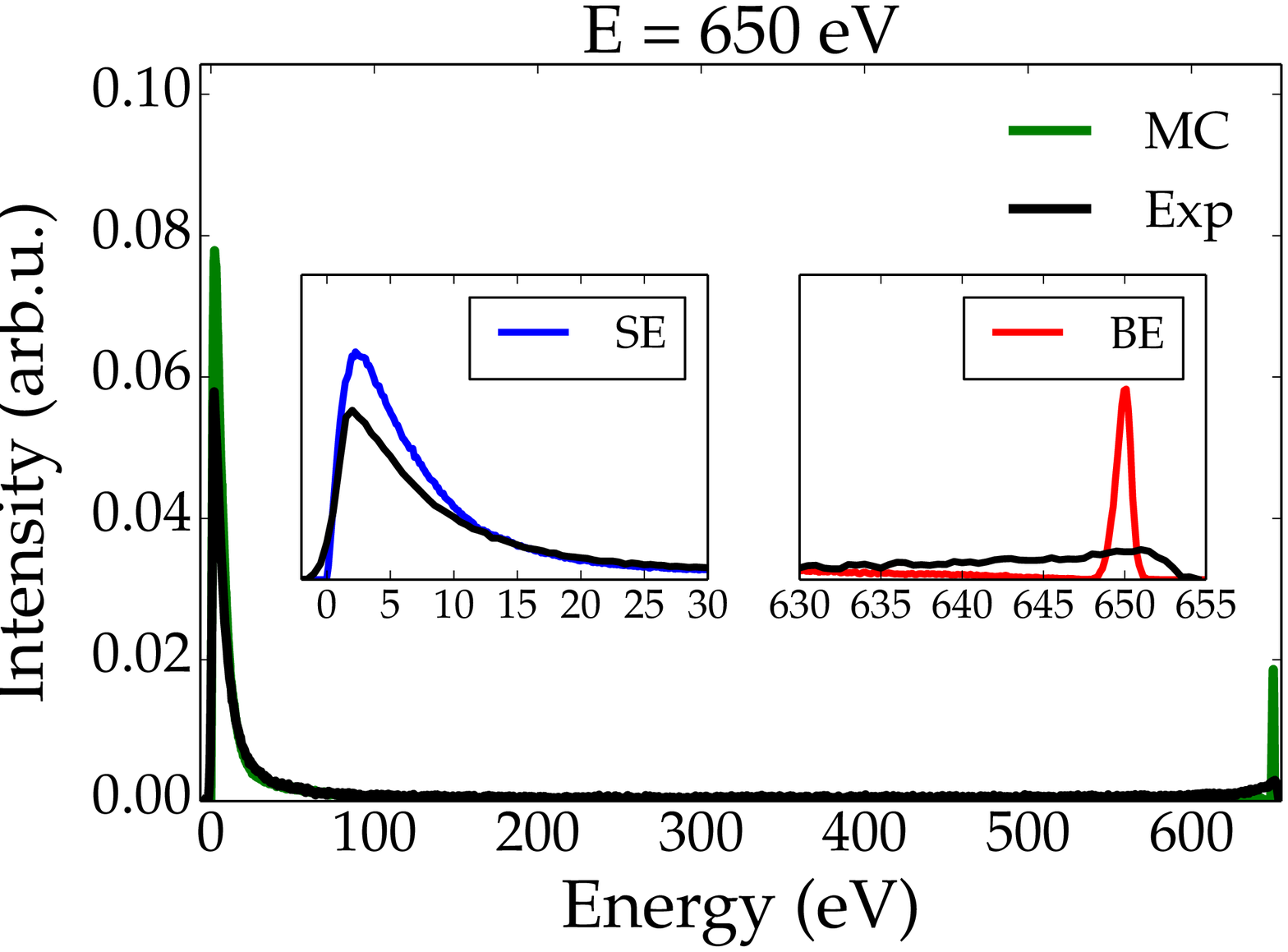}
    \includegraphics[width = 0.40\textwidth]{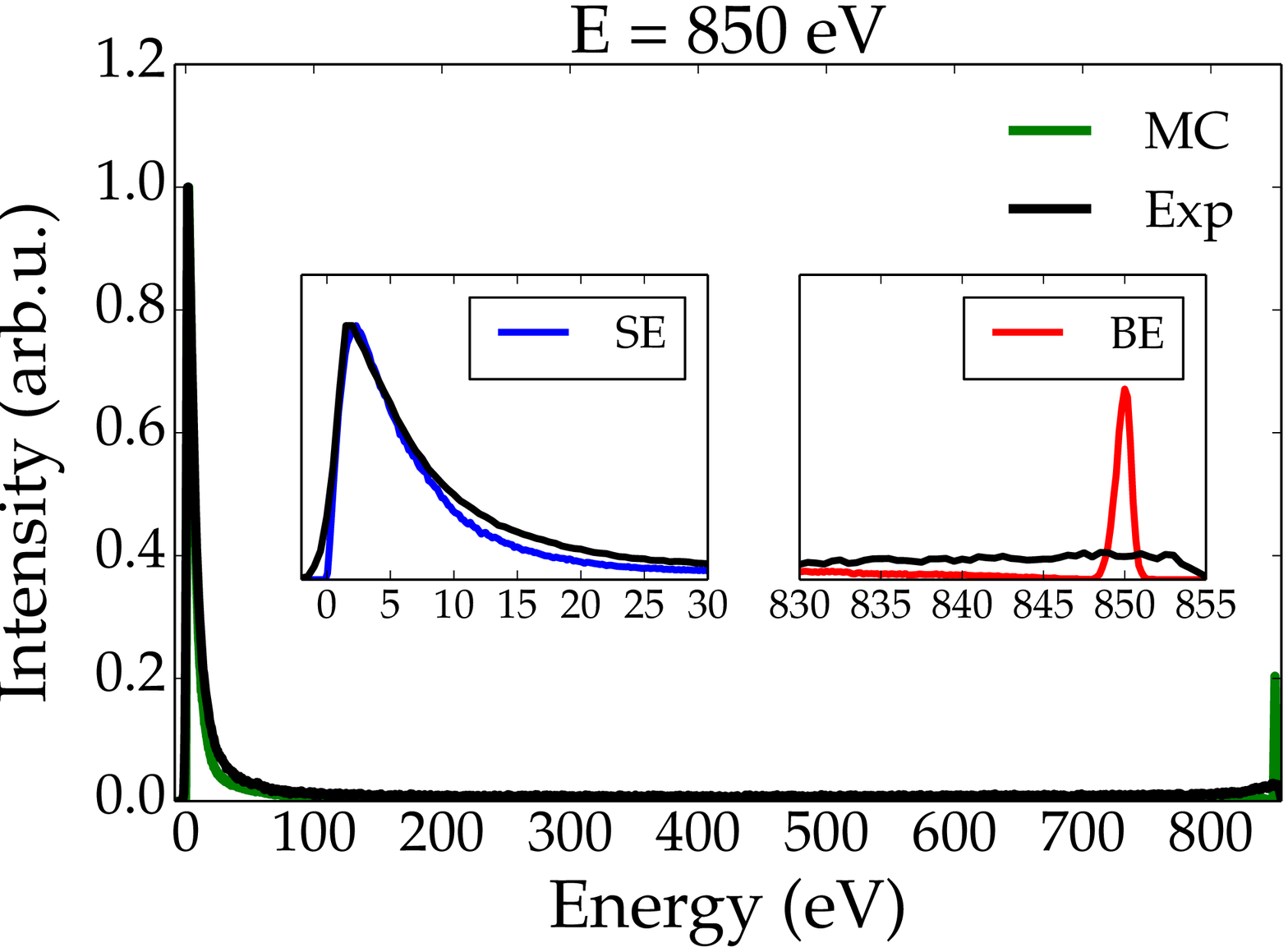}
    \includegraphics[width = 0.40\textwidth]{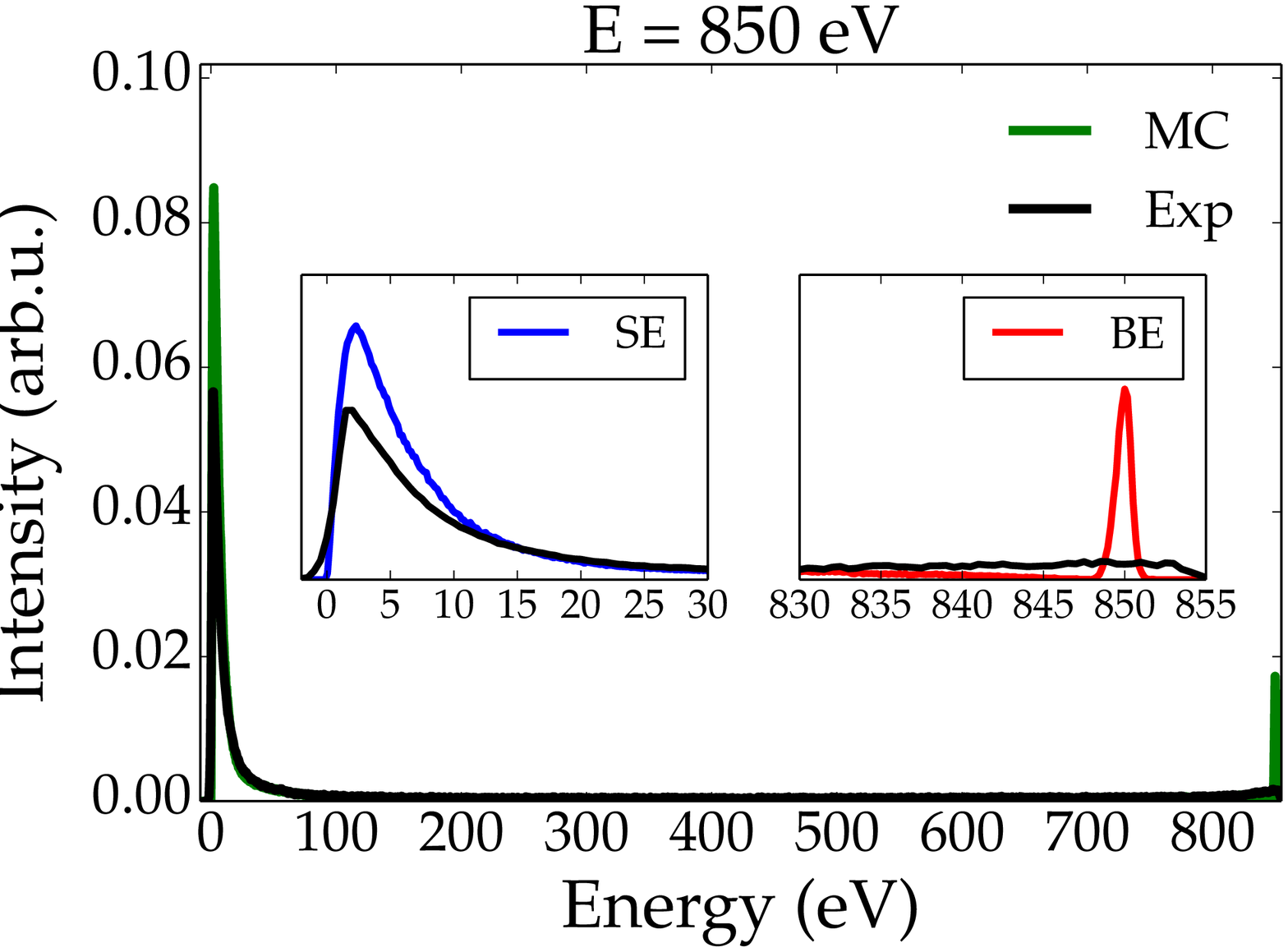}\\
    \caption{Electron energy emission spectra for different initial kinetic energies. The panels on the left side of the figure report spectra normalized at a common height of the secondary electron emission peak, while the panels on the right show spectra normalized at a common total area of the spectrum. The insets in each panel display the two main contributions to the spectra, that is the secondary electron (SE) emission peak (blue curve) and the back-scattered electrons (BE, red curve), respectively. The experimental data are shown as black lines}
    \label{fig:reel1}
\end{figure*}
As it is reported above, we notice that the experimental spectra were acquired  
with a (RF) analyzer which is known to cause a  
characteristic broadening of the elastic peak due to poor   resolution  at high energies and a strong  asymmetry on  the low energy side due to the integration of the background \cite{gergely2002elastic, sulyok1992spectrometer}. 
\begin{figure*}[hbt!]
    \centering
\includegraphics[width = 0.40\textwidth]{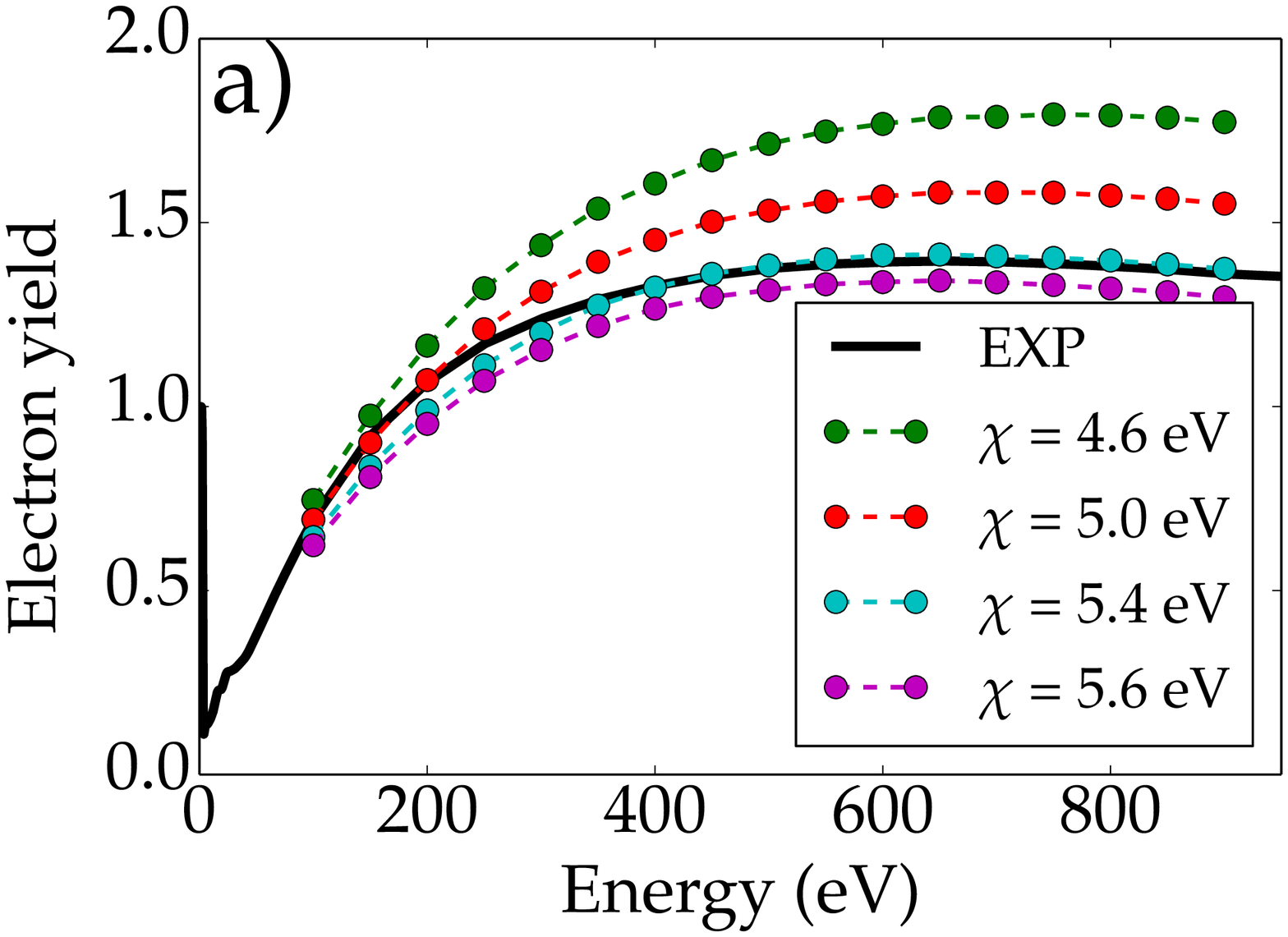}
    \includegraphics[width = 0.40\textwidth]{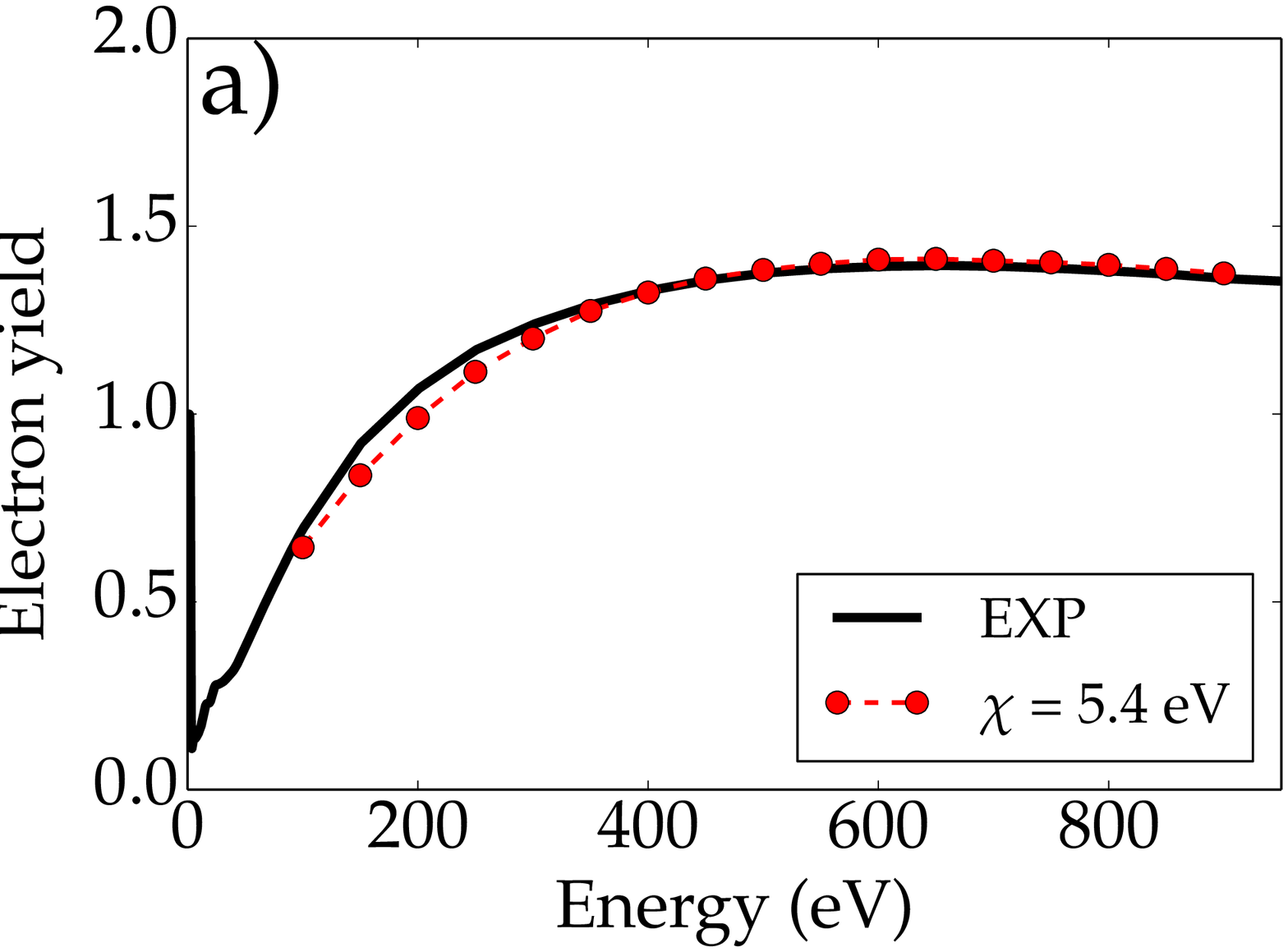}\\
    \includegraphics[width = 0.40\textwidth]{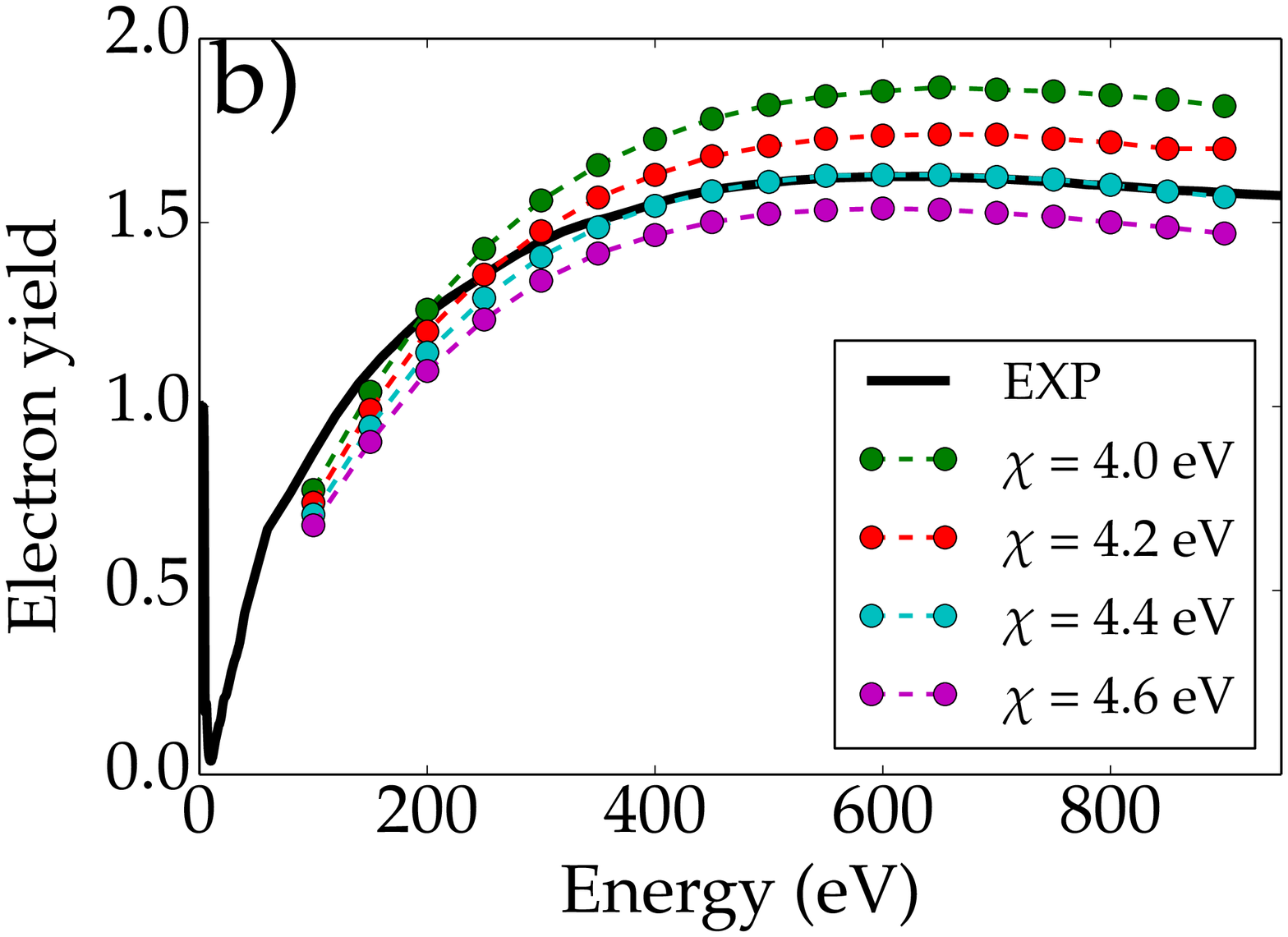}
    \includegraphics[width = 0.40\textwidth]{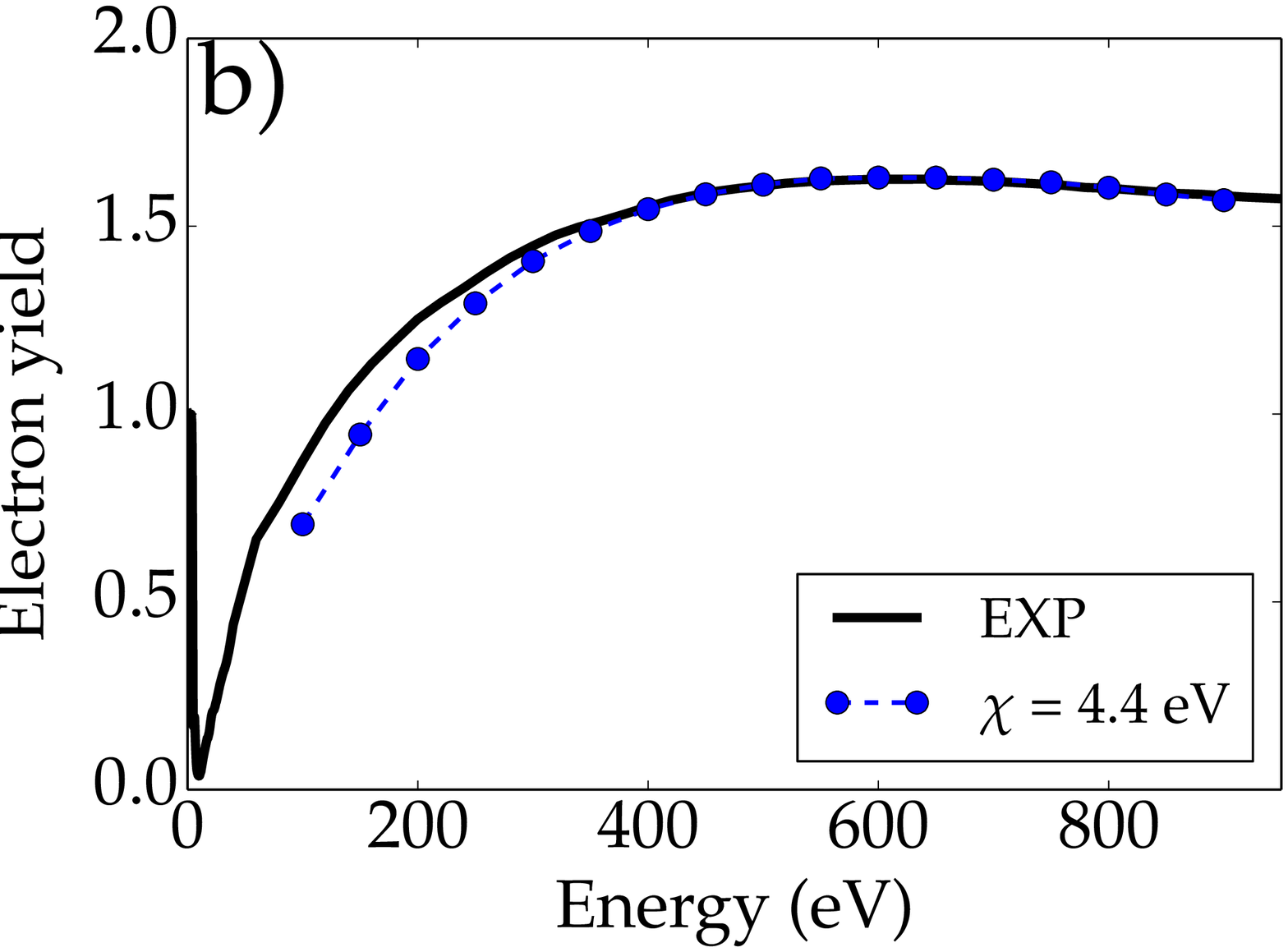}\\
    \includegraphics[width = 0.40\textwidth]{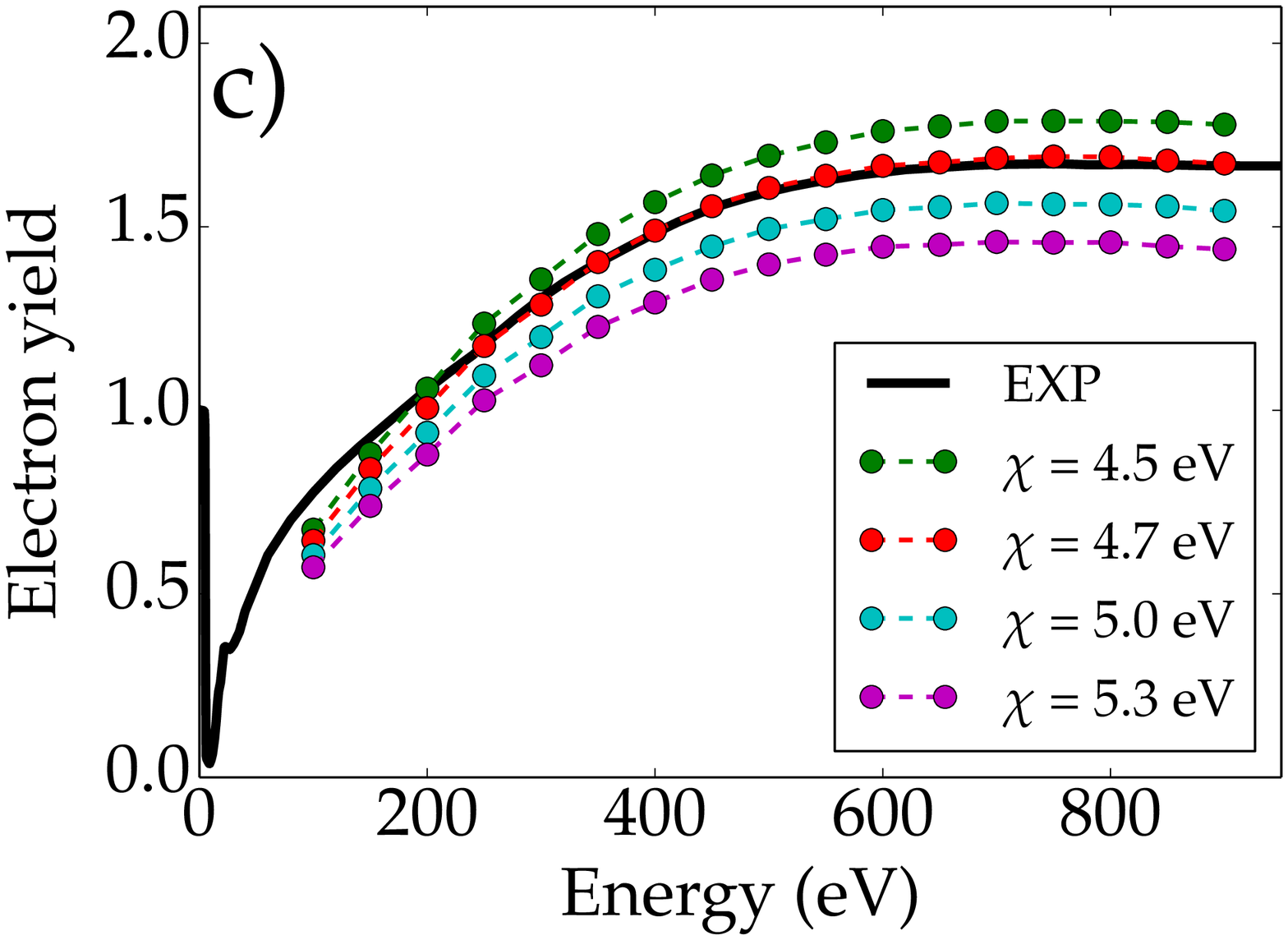}
    \includegraphics[width = 0.40\textwidth]{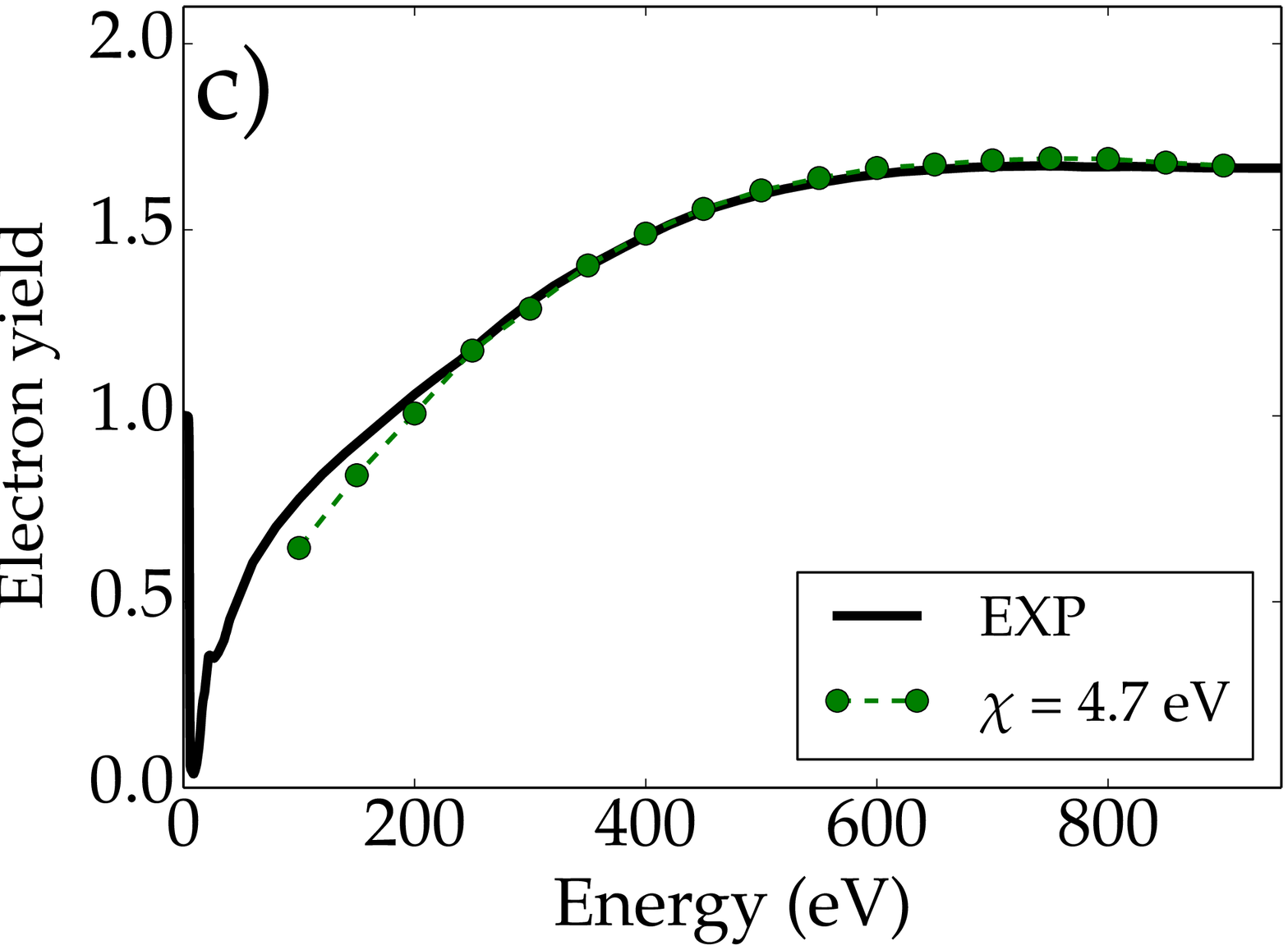}\\
    \caption{Electron yield curves of a) Cu, b) Ag, and c) Au as a function of the initial primary electron beam kinetic energy, for different values of the work functions $\chi$ (left panels). In black we report the experimental data \citep{gonzalez2017secondary}. In the right panels we report the yield spectra for the value of the work function leading to the best agreement between simulations and experiments}
    \label{fig:yield}
\end{figure*}
Thus, the integrated area of the whole elastic  peak is always higher than that obtained by our MC simulations \cite{zeze1997retarding, gruzza2010monte}.
This is the reason why the MC secondary electron peaks
are more intense than the experimental ones when the spectra are normalized at a common area (see the panels at the right side of Fig. 9).
Nevertheless, by
normalizing to a common height of the secondary electron emission peak, a remarkable agreement is obtained between the secondary electrons MC and experimental lineshapes.

\subsection{Electron yield}

The electron yield is defined as the total number of emitted electrons divided by the number of electrons in the beam. This quantity was calculated in the case of Cu, Ag and Au for different initial kinetic energies. To calculate the electron yield spectra, we set to 10$^6$ the number of electrons in the beam. 
In Fig. \ref{fig:yield} we compare the theoretical electron yield curves with the experimental data by Gonzales {\it et al.} \citep{gonzalez2017secondary}. In the left hand side of Fig. 10 we tested the dependence of the yield upon a reasonable change of the work function $\chi$: thus, different simulations were carried out by changing this parameter.\\
\indent These results show that spectra calculated with a higher value of $\chi$ display lower intensities than those obtained with a lower $\chi$. Indeed, it makes sense that an increase of the emission energy barrier, that is the work function, results in a decreasing number of electrons emerging from the surface.
The best agreement with the experimental data \citep{gonzalez2017secondary} is obtained for $\chi=5.4$ eV for copper (experimental value is 4.6 eV \citep{gonzalez2017secondary}), $\chi=4.4$ eV for silver (experimental value is 4.4 eV \citep{gonzalez2017secondary}), and $\chi=4.7$ eV for gold (experimental value is 5.3 eV \citep{gonzalez2017secondary}) respectively. We notice that for silver we are able to reproduce the yield experimental spectra with a exceptional agreement of the work function between simulations and measurements. In the case of copper and gold this agreement is anyway rather good. 
 
\section{Conclusions}

In this work a Monte Carlo approach developed for modeling the electron transport in metallic samples was described. Our approach is capable to deliver the accurate calculation of the secondary electron emission spectra and of the electron yield curves. We have simulated these characteristics for three different metals, that is copper, silver and gold. MC simulations were carried out also to analyse how tuning the work functions of the metals may affect the secondary electron yield. We found out that a remarkable agreement between simulated and experimental yield spectra could be reached by setting the values of the work functions for the different metals very close to those obtained experimentally \citep{gonzalez2017secondary}.
As a further improvement of our MC code suite, the possibility to model a tailored target surface morphology will be introduced, in order to investigate the effect that a given shape can have \citep{ye2017mechanism} in increasing or decreasing the electron yield according to the needed application.

\newpage
\section{Acknowledgements}
N.M.P. is supported by the European Commission under the Graphene Flagship Core 2 grant No. 
785219 (WP14, "Composites") and FET Proactive ("Neurofibres") grant No. 732344 as well as by the Italian Ministry of Education, University and Research (MIUR) under the ‘‘Departments of Excellence’’ grant L.232/2016. The authors gratefully acknowledge the Gauss Centre for Supercomputing for funding this project by providing computing time on the GCS Supercomputer JUQUEEN at J\"ulich Supercomputing Centre (JSC) \citep{juqueen}. Furthermore, the authors acknowledge FBK for providing unlimited access
to the KORE computing facility.

\bibliographystyle{elsarticle-num}


\begin{thebibliography}{10}
\expandafter\ifx\csname url\endcsname\relax
  \def\url#1{\texttt{#1}}\fi
\expandafter\ifx\csname urlprefix\endcsname\relax\def\urlprefix{URL }\fi
\expandafter\ifx\csname href\endcsname\relax
  \def\href#1#2{#2} \def\path#1{#1}\fi

\bibitem{seiler1983secondary}
H.~Seiler, Secondary electron emission in the scanning electron microscope,
  Journal of Applied Physics 54~(11) (1983) R1--R18.

\bibitem{goldstein2017scanning}
J.~I. Goldstein, D.~E. Newbury, J.~R. Michael, N.~W. Ritchie, J.~H.~J. Scott,
  D.~C. Joy, Scanning electron microscopy and X-ray microanalysis, Springer,
  2017.

\bibitem{sauli1997gem}
F.~Sauli, Gem: A new concept for electron amplification in gas detectors,
  Nuclear Instruments and Methods in Physics Research Section A: Accelerators,
  Spectrometers, Detectors and Associated Equipment 386~(2-3) (1997) 531--534.

\bibitem{benlloch1998further}
J.~Benlloch, A.~Bressan, M.~Cape{\'a}ns, M.~Gruw{\'e}, M.~Hoch, J.~Labb{\'e},
  A.~Placci, L.~Ropelewski, F.~Sauli, Further developments and beam tests of
  the gas electron multiplier (gem), Nuclear Instruments and Methods in Physics
  Research Section A: Accelerators, Spectrometers, Detectors and Associated
  Equipment 419~(2-3) (1998) 410--417.

\bibitem{lyneis1977elimination}
C.~Lyneis, H.~Schwettman, J.~Turneaure, Elimination on electron multipacting in
  superconducting structures for electron accelerators, Applied Physics Letters
  31~(8) (1977) 541--543.

\bibitem{somersalo1998computational}
E.~Somersalo, D.~Proch, P.~Yl{\"a}-Oijala, J.~Sarvas, Computational methods for
  analyzing electron multipacting in rf structures, Part. Accel. 59 (1998)
  107--141.

\bibitem{cimino2004can}
R.~Cimino, I.~Collins, M.~Furman, M.~Pivi, F.~Ruggiero, G.~Rumolo,
  F.~Zimmermann, Can low-energy electrons affect high-energy physics
  accelerators?, Physical review letters 93~(1) (2004) 014801.

\bibitem{cimino2012nature}
R.~Cimino, M.~Commisso, D.~Grosso, T.~Demma, V.~Baglin, R.~Flammini,
  R.~Larciprete, Nature of the decrease of the secondary-electron yield by
  electron bombardment and its energy dependence, Physical review letters
  109~(6) (2012) 064801.

\bibitem{larciprete2013secondary}
R.~Larciprete, D.~Grosso, M.~Commisso, R.~Flammini, R.~Cimino, Secondary
  electron yield of cu technical surfaces: Dependence on electron irradiation,
  Physical Review Special Topics-Accelerators and Beams 16~(1) (2013) 011002.

\bibitem{cimino2014electron}
R.~Cimino, T.~Demma, Electron cloud in accelerators, International Journal of
  Modern Physics A 29~(17) (2014) 1430023.

\bibitem{baglin2000secondary}
V.~Baglin, J.~Bojko, C.~Scheuerlein, O.~Gr{\"o}bner, M.~Taborelli, B.~Henrist,
  N.~Hilleret, The secondary electron yield of technical materials and its
  variation with surface treatments, Tech. rep. (2000).

\bibitem{vallgren2011amorphous}
C.~Y. Vallgren, G.~Arduini, J.~Bauche, S.~Calatroni, P.~Chiggiato, K.~Cornelis,
  P.~C. Pinto, B.~Henrist, E.~M{\'e}tral, H.~Neupert, et~al., Amorphous carbon
  coatings for the mitigation of electron cloud in the cern super proton
  synchrotron, Physical Review Special Topics-Accelerators and Beams 14~(7)
  (2011) 071001.

\bibitem{montero2014secondary}
I.~Montero, L.~Aguilera, M.~E. D{\'a}vila, V.~C. Nistor, L.~A. Gonz{\'a}lez,
  L.~Gal{\'a}n, D.~Raboso, R.~Ferritto, Secondary electron emission under
  electron bombardment from graphene nanoplatelets, Applied Surface Science 291
  (2014) 74--77.

\bibitem{valizadeh2014low}
R.~Valizadeh, O.~B. Malyshev, S.~Wang, S.~A. Zolotovskaya, W.~Allan~Gillespie,
  A.~Abdolvand, Low secondary electron yield engineered surface for electron
  cloud mitigation, Applied Physics Letters 105~(23) (2014) 231605.

\bibitem{valizadeh2017reduction}
R.~Valizadeh, O.~Malyshev, S.~Wang, T.~Sian, M.~D. Cropper, N.~Sykes, Reduction
  of secondary electron yield for e-cloud mitigation by laser ablation surface
  engineering, Applied Surface Science 404 (2017) 370--379.

\bibitem{calatroni2017first}
S.~Calatroni, E.~G.-T. Valdivieso, H.~Neupert, V.~Nistor, A.~T.~P. Fontenla,
  M.~Taborelli, P.~Chiggiato, O.~Malyshev, R.~Valizadeh, S.~Wackerow, et~al.,
  First accelerator test of vacuum components with laser-engineered surfaces
  for electron-cloud mitigation, Physical Review Accelerators and Beams 20~(11)
  (2017) 113201.

\bibitem{vaughan1989new}
J.~R.~M. Vaughan, A new formula for secondary emission yield, IEEE Transactions
  on Electron Devices 36~(9) (1989) 1963--1967.

\bibitem{dionne1975origin}
G.~F. Dionne, Origin of secondary-electron-emission yield-curve parameters,
  Journal of Applied Physics 46~(8) (1975) 3347--3351.

\bibitem{shih1997secondary}
A.~Shih, J.~Yater, C.~Hor, R.~Abrams, Secondary electron emission studies,
  Applied surface science 111 (1997) 251--258.

\bibitem{lin2005new}
Y.~Lin, D.~C. Joy, A new examination of secondary electron yield data, Surface
  and interface analysis 37~(11) (2005) 895--900.

\bibitem{xie2009formula}
A.-G. Xie, H.-F. Zhao, B.~Song, Y.-J. Pei, The formula for the secondary
  electron yield at high incident electron energy from silver and copper,
  Nuclear Instruments and Methods in Physics Research Section B: Beam
  Interactions with Materials and Atoms 267~(10) (2009) 1761--1763.

\bibitem{dapor2011secondary}
M.~Dapor, Secondary electron emission yield calculation performed using two
  different monte carlo strategies, Nuclear Instruments and Methods in Physics
  Research Section B: Beam Interactions with Materials and Atoms 269~(14)
  (2011) 1668--1671.

\bibitem{mott1929scattering}
N.~Mott, The scattering of fast electrons by atomic nuclei, Proc. R. Soc.
  London, Ser. A 124~(794) (1929) 425.

\bibitem{Ritchie_PhysRev_1957}
R.~H. Ritchie, Plasma losses by fast electrons in thin films, Phys. Rev. 106
  (1957) 874.

\bibitem{werner2009optical}
W.~S. Werner, K.~Glantschnig, C.~Ambrosch-Draxl, Optical constants and
  inelastic electron-scattering data for 17 elemental metals, Journal of
  Physical and Chemical Reference Data 38~(4) (2009) 1013--1092.

\bibitem{cimino2015detailed}
R.~Cimino, L.~A. Gonzalez, R.~Larciprete, A.~Di~Gaspare, G.~Iadarola,
  G.~Rumolo, Detailed investigation of the low energy secondary electron yield
  of technical cu and its relevance for the lhc, Physical Review Special
  Topics-Accelerators and Beams 18~(5) (2015) 051002.

\bibitem{gonzalez2017secondary}
L.~Gonzalez, M.~Angelucci, R.~Larciprete, R.~Cimino, The secondary electron
  yield of noble metal surfaces, AIP Advances 7~(11) (2017) 115203.

\bibitem{gergely2002elastic}
G.~Gergely, Elastic backscattering of electrons: determination of physical
  parameters of electron transport processes by elastic peak electron
  spectroscopy, Progress in surface science 71~(1-4) (2002) 31--88.

\bibitem{sulyok1992spectrometer}
A.~Sulyok, G.~Gergely, B.~Gruzza, Spectrometer corrections for a retarding
  field analyser used for elastic peak electron spectroscopy and auger electron
  spectroscopy, Acta Physica Hungarica 72~(1) (1992) 107.

\bibitem{Dapor_book_blu}
M.~Dapor, Transport of Energetic Electrons in Solids, Springer Tracts in Modern
  Physics, 2017.

\bibitem{jablonski2004comparison}
A.~Jablonski, F.~Salvat, C.~J. Powell, Comparison of electron
  elastic-scattering cross sections calculated from two commonly used atomic
  potentials, Journal of physical and chemical reference data 33~(2) (2004)
  409--451.

\bibitem{PhysRevA.36.467}
F.~Salvat, J.~Martinez, R.~Mayol, J.~Parellada, Analytical
  {D}irac-{H}artree-{F}ock-{S}later screening function for atoms
  ({Z}=1\char21{}92), Phys. Rev. A 36 (1987) 467--474.

\bibitem{jablonski2010nist}
A.~Jablonski, F.~Salvat, C.~Powell, Nist electron elastic-scattering
  cross-section database, NIST Standard Reference Database 64.

\bibitem{AZZOLINI2017299}
M.~Azzolini, T.~Morresi, G.~Garberoglio, L.~Calliari, N.~M. Pugno, S.~Taioli,
  M.~Dapor,
  \href{http://www.sciencedirect.com/science/article/pii/S0008622317302816}{Monte
  carlo simulations of measured electron energy-loss spectra of diamond and
  graphite: Role of dielectric-response models}, Carbon 118 (2017) 299 -- 309.
\newblock \href
  {http://dx.doi.org/https://doi.org/10.1016/j.carbon.2017.03.041}
  {\path{doi:https://doi.org/10.1016/j.carbon.2017.03.041}}.
\newline\urlprefix\url{http://www.sciencedirect.com/science/article/pii/S0008622317302816}

\bibitem{doi:10.1002/pssb.200982339}
S.~Taioli, P.~Umari, M.~M. De~Souza,
  \href{https://onlinelibrary.wiley.com/doi/abs/10.1002/pssb.200982339}{Electronic
  properties of extended graphene nanomaterials from gw calculations}, physica
  status solidi (b) 246~(11‐12)  2572--2576.
\newblock \href
  {http://arxiv.org/abs/https://onlinelibrary.wiley.com/doi/pdf/10.1002/pssb.200982339}
  {\path{arXiv:https://onlinelibrary.wiley.com/doi/pdf/10.1002/pssb.200982339}},
  \href {http://dx.doi.org/10.1002/pssb.200982339}
  {\path{doi:10.1002/pssb.200982339}}.
\newline\urlprefix\url{https://onlinelibrary.wiley.com/doi/abs/10.1002/pssb.200982339}

\bibitem{Tanuma_SIA_2009}
S.~Tanuma, C.~Powell, D.~Penn, Calculations of electron inelastic mean free
  paths. {IX}. {D}ata for 41 elemental solids over the 50 e{V} to 30 ke{V}
  range, Surface and Interface Analysis 43 (2011) 689.

\bibitem{montanari2007calculation}
C.~Montanari, J.~Miraglia, S.~Heredia-Avalos, R.~Garcia-Molina, I.~Abril,
  Calculation of energy-loss straggling of {C}, {Al}, {Si}, and {Cu}for fast
  {H}, {He}, and {L}i ions, Physical Review A 75~(2) (2007) 022903.

\bibitem{denton2008influence}
C.~Denton, I.~Abril, J.~Garcia-Molina, Rand Moreno-Mar{\'\i}n,
  S.~Heredia-Avalos, Influence of the description of the target energy-loss
  function on the energy loss of swift projectiles, Surface and Interface
  Analysis 40~(11) (2008) 1481--1487.

\bibitem{zeze1997retarding}
D.~Zeze, L.~Bideux, B.~Gruzza, F.~Go{\l}ek, D.~Da{\'n}ko, S.~Mroz, Retarding
  field analyser used in elastic peak electron spectroscopy, Vacuum 48~(3-4)
  (1997) 399--401.

\bibitem{gruzza2010monte}
B.~Gruzza, S.~Chelda, C.~Robert-Goumet, L.~Bideux, G.~Monier, Monte carlo
  simulation for multi-mode elastic peak electron spectroscopy of crystalline
  materials: Effects of surface structure and excitation, Surface Science
  604~(2) (2010) 217--226.

\bibitem{ye2017mechanism}
M.~Ye, D.~Wang, Y.~He, Mechanism of total electron emission yield reduction
  using a micro-porous surface, Journal of Applied Physics 121~(12) (2017)
  124901.

\bibitem{juqueen}
{J\"{u}lich Supercomputing Centre},
  \href{http://dx.doi.org/10.17815/jlsrf-1-18}{{JUQUEEN: IBM Blue Gene/Q
  Supercomputer System at the J\"{u}lich Supercomputing Centre}}, Journal of
  large-scale research facilities 1~(A1).
\newblock \href {http://dx.doi.org/10.17815/jlsrf-1-18}
  {\path{doi:10.17815/jlsrf-1-18}}.
\newline\urlprefix\url{http://dx.doi.org/10.17815/jlsrf-1-18}

\end{thebibliography}

\end{document}